\documentclass[11pt,a4paper]{article}
\pdfoutput=1
\usepackage{jheppub}
\usepackage{amsmath,amstext,amssymb}
\usepackage{graphicx}
\usepackage{feynmp}
\usepackage{epstopdf}
\usepackage{caption}
\usepackage{float}

\def\be{\begin{equation}}
\def\ee{\end{equation}}
\def\ba{\begin{eqnarray}}
\def\ea{\end{eqnarray}}
\newcommand{\psidag}{\psi^{\dag}}
\newcommand{\phidag}{\phi^{\dag}}

\newcommand{\tildepsi}{\tilde{\psi}}
\newcommand{\tildepsidag}{\tilde{\psi}^{\dag}}

\newcommand{\bq}{ \textbf{q}}

\newcommand{\bbeta}{\textbf{B}}

\usepackage{graphics}
\usepackage{graphicx}% Include figure files
\usepackage{dcolumn}% Align table columns on decimal point
\usepackage{bm}% bold math
\usepackage{epsfig}
\usepackage{graphicx}
\usepackage{multirow}
\usepackage{dcolumn}% Align table columns on decimal point
\usepackage{graphicx,epsfig}%

\begin{document}

%\preprint{preprint USM-TH-298}

\title{\center Could reggeon field theory be an effective theory for QCD in the Regge limit?}

\author[a]{Jochen Bartels,}%,\note{Corresponding author.}}
\author[b]{Carlos Contreras,}
\author[c]{G. P. Vacca}%\note{Also at Some University.}}

\affiliation[a]{II. Institut f\"{u}r Theoretische Physik, Universit\"{a}t Hamburg, Luruper Chaussee 149,\\
D-22761 Hamburg, Germany}
\affiliation[b]{Departamento de Fisica, Universidad Tecnica Federico Santa Maria, Avda.España 1680, Casilla 110-V, Valparaiso, Chile }
\affiliation[c]{INFN Sezione di Bologna, DIFA, via Irnerio 46, I-40126 Bologna, Italy}

% e-mail addresses: one for each author, in the same order as the authors
\emailAdd{jochen.bartels@desy.de}
\emailAdd{carlos.contreras@usm.cl}
\emailAdd{vacca@bo.infn.it}

%%%%%%%%%%%%%%%%%%%%%%
\abstract{In this paper we investigate the possibility whether, in the extreme limit of high energies and large transverse distances, 
reggeon field theory might serve as an effective theory of high energy scattering for strong interactions. 
We analyse the functional renormalization group equations (flow equations) of reggeon field theory and search for   
fixed points in the space of (local) reggeon field theories. 
We study in complementary ways the candidate for the scaling solution, investigate its main properties and
briefly discuss possible physical interpretations.}     
\date{\today}
%%%%%%%%%%%%%%%%%%%%%%

%\pacs{12.38.Cy, 12.38.Aw,12.40.Vv}

\maketitle
 
%%%%%%%%%%%%%%%%%%%%%%%%%%%%%%%%
\section{Introduction}
%%%%%%%%%%%%%%%%%%%%%%%%%%%%%%%%
\label{sec:intr}
To find a bridge between short and long distances physics in the Regge limit of high energy QCD remains 
a topic of high interest. At small transverse distances where perturbation theory can be applied QCD predicts the BFKL Pomeron~\cite{BFKL} with an intercept bigger than unity and a very small Pomeron slope.  
This BFKL Pomeron repreresents the basic building block of QCD reggeon field  theory \cite{Lipatov:1995pn}in which the reggeized gluon 
is the basic field and the Pomeron is generated as a composite state of two or more reggeized gluons. 
Within this perturbative QCD reggeon field theory, the BFKL Pomeron has infinitely many (nonlocal) selfcouplings, 
\cite{Bartels:1994jj} leading to Pomeron loops  and to higher bound states composed of $4, 6, ...$ reggeized gluons, with a simplified picture in the large $N_c$ limit~\cite{Braun:1997nu, Bartels:2004hb}.
On the other hand, high energy scattering of hadrons strongly depends upon large transverse distances where perturbation theory is not applicable. As the most promising 
theoretical concept, again Regge theory ~\cite{Gribov:1968fg,AB,Migdal:1973gz,Sugar:1974td,Moshe:1977fe} can be used, but the parameters have to be taken from data. 
Most prominent and phenomenologically successful examples include the Regge pole model of Donnachie and Landshoff \cite{Donnachie:2013xia}, 
with a Pomeron intercept slightly above one (but smaller than the BFKL intercept),  the reggeon field theory model of Kaidalov et al \cite{Kaidalov:1986tu}, 
and, more recently, the two models based upon summation of Pomeron diagrams of the Tel-Aviv \cite{Gotsman:2008tr} and the Durham group \cite{Ryskin:2011qe}. 
Most important, the observed growth of the total cross section of $pp$ scattering can be described by a Pomeron pole 
(and corrections due to a two-Pomeron cut) with intercept slightly above one and a nonzero Pomeron slope. 

We therefore see evidence that in both regions  - short and long transverse distances - we have the same structure: reggeon field (RFT) which lives in one time (rapidity)
and two space dimensions (transverse distances) (these variables are conjugate to reggeon energy $\omega$ and and transverse momenta $k$) \cite{Gribov:1968fc}:
\be
S(\psi,\psidag) = \int d\tau d^2x {\cal{L}}(\psi(\tau,x) \psidag(\tau,x)).
\ee
In its simplest version, RFT is based upon Regge poles. What is different in the two regions are the parameters of the Regge poles (intercepts, slopes) and their interactions. 
As we have already said before, at short distances QCD perturbation theory provides information,  whereas the parameters describing the long distance region, so far, have to taken from data. 
The theoretical challenge then is to find a bridge between the two regions.

In this paper we present  a first step in searching for a  connection between the perturbative UV region with 
the nonperturbative IR region. We want to address the following question: can RFT be considered as useful effective description in the large distance region, 
which eventually can be linked to what we know from perturbative high energy QCD in the region of small transverse distances (ultraviolet region)? As a start  we consider 
a class of local reggeon field theories with all general interactions. In a pertubative language, we include not only the triple Pomeron vertex but also quartic and higher couplings.
For our investigation we make use of nonperturbative renormalization group techniques,  investigate the fixed point structure and study features of the flow,
which shed light on some universal features of the theory.
% in the region of long distances and high energies (infrared region). 
%In this first paper, we will concentrate on the longe distance/infrared limit, and we will  not yet address the flow from the UV region to IR region, Nevertheless 

It may be useful to recapitulate a few results obtained in the short distance region.
Starting from the BFKL Pomeron, QCD reggeon field theory has been established as the 
field theory of interacting reggeizing gluons \cite{Lipatov:1995pn} where the BFKL Pomeron appears as the bound state of two gluons. 
In the leading approximation the BFKL Pomeron is scale invariant and in the complex angular momentum plane generates a fixed cut singularity above one 
which does not depend upon the momentum transfer. If one imposes  boundary conditions in the IR region, the BFKL Pomeron turns into a sequence of Regge poles \cite{Lipatov:1985uk}, 
some of which have intercepts greater than one. It is expected that these poles have small $t$-slopes. The triple Pomeron vertex 
has been derived \cite{Bartels:1994jj,Braun:1997nu} from the $2\to4$ gluon transition vertex. It is important to note that 
the short distance Pomeron by itself is a bilocal field, and the reggeon field theory of interacting Pomerons is nonlocal.  
All this suggests that the flow which we will have to investigate in future steps
will start, in the UV region, with an intercept above unity and a small but nonzero Pomeron slope.     

As to the long distance region, studies of field theory models of interacting Pomerons have been started by Gribov  \cite{Gribov:1968fg,Gribov:1968uy} many years ago. An important step has been taken in \cite{AB,Migdal:1973gz,Sugar:1974td} where reggeon field theory with zero renormalized reggeon mass has been investigated by means of the Callan-Symanzik equation and the $\epsilon$ 
expansion in the vicinity of four transverse dimension. The key result was the existence of an infrared fixpoint which leads to scaling laws for the Pomeron Green's function.
Subsequently numerous studies of reggeon field theory with zero transverse dimensions have been performed in which the Pomeron intercept was allowed to vary.  
Our analysis aims at a wide class of reggeon field theories; in particular we do not impose any constraint on the Pomeron mass,
and we work in $d=2$ transverse dimensions including truly non perturbative contributions.
We expect that the results obtained earlier should be identified as particular cases of this more general approach.  

Our main tool of investigating RFT, and in particular its fixed point structure under RG flow,
is the functional renormalization group approach ~\cite{Wetterich, Morris:1994ie, Gies:2006wv}, for the generator of the proper vertices of the theory also called effective average action (EAA),
which has successfully been applied to numerous problems in statistical mechanics, in particle physics, and in quantum gravity. 
In short, in this approach we study the effective action of a sequence RFT's as a function of an infrared regulator $k>0$ and search for fixed points of the flow. 
The dependence on $k$ is captured by the flow equations which have to be solved by suitable approximations. 
One main result will be the existence of a fixed point with one relevant direction: we will analyse this fixed point and  the effective potential. We also will present first indications of the possible physical significance of this fixed pont.        

This work is organized as follows. We first describe the general setup, We then present results of our fixed point analysis, and we describe the effective potential at the fixed point. 
In our final part we compute trajectories of physical parameters (Pomeron intercept and  Pomeron interaction vertices) and derive first hints at a physical interpretation.  

%%%%%%%%%%%%%%%%%%%%%%%%%%%%%%%%%%%%%%%%%%%%%%%%
\section{The general setup}
%%%%%%%%%%%%%%%%%%%%%%%%%%%%%%%%%%%%%%%%%%%%%%%%
\subsection{Introductory remarks} 
Before starting our investigation let us make some general considerations.
As already anticipated the tool we are going to use is the renormalisation group technique, in particular we shall study the flow equation 
of the EAA. This equation describes the change of the generator of the proper vertices of the theory, $\Gamma_k[\phi]$, 
as a function of the infrared cutoff $k$ which controls the range of modes which are integrated out. 
For $k>0$ the infrared region is regulated by some cutoff operator $R_k$ which usually is associated to a quadratic form. 
In the limit $k\to 0$ one finds the full effective action of the theory.

In general the EAA is a non local functional and cannot be written in terms of a local lagrangian, exactly as its standard effective action counterpart. 
Since it is impossible to deal with such a problem exactly, one looks for some simplifications: 
indeed many properties of the dynamical flow can be studied choosing a truncation which consists in projecting the generating functional onto a subspace. 
One of the most popular ones is based on a derivative expansion: it starts from the local potential approximation (LPA), 
where - apart from a simple kinetic term describing the propagation of the fields-  one allows for a pure local potential term  $V$ in the lagrangian.
The full leading order (LO) in the derivative expansion contains arbitrary field functions entering the two derivatives (kinetic) terms.
To study, in approximate way, the anomalous dimensions one can employ a intermediated approximation scheme (LPA'),
where the kinetic terms are multiplied by scale-dependent constants $Z_k$ which are independent of the fields. This truncation scheme is what we shall use in our investigation.
Even this simple approximation includes an infinite number of couplings which are the coordinates in a basis of ultralocal operators.
Often the potential is expanded in power series of the fields. We stress that this may limited if there is a finite radius of convergence.
The RG flow equation allows to study two important features of the dynamics: 

- first one may ask which are the functionals which are invariant under the flow, the fixed point (FP) of the flow,
and how the flow behaves close to it (at linear level the spectral analysis around a FP leads to the critical exponents predicting a universal behavior).
This analysis deals with the critical behavior and has to be done in terms of dimensionless quantities,
keeping in mind that many physical aspects are determined by the dimensionful ones.

- second one can study the flow from some bare condition in the UV regime towards the IR one
and investigate the approximated form of the effective action, depending on the different regions (phases) one starts from. 

We note that the fixed points can be interpreted as either UV or IR, and that there might exist a particular flow from UV to IR which connects two different fixed points.
One may face two interesting situations, related to the concept of emergence of an effective theory from a microscopic "more" fundamental one, after having considered a change of description of the degrees of freedom (fields):

1) If the bare action of the effective theory is located on the critical surface of a FP, spanned by the irrelevant (UV repulsive and IR attractive) directions,
the flow will fall into the FP and  a critical theory wikk be reached.

2) The bare action is out of the critical surface. If there exist a UV FP with a finite number of relevant directions 
(eigenvectors of the associated linearized flow with positive critical exponents,  i.e. UV attractive, are othogonal to the critical surface), 
a flow starting from a point close to such a FP towards the IR will tend to span a finite dimensional subspace.
Let us remind that if the bare action is exactly on the submanifold attracted to the FP when flowing to the UV, then the theory is said to be renormalizable (in the general asymptotically safe sense), one can safely remove any UV cutoff maintaining finite dimensionless quantities and only a finite number of coupling is independent leading to full predictivity with a finite number of measurements.

Starting from the fields of the fundamental degrees of freedom (d.o.f.) one may see that a convenient description arises performing a change of field variables, with eventually new symmetries for the new d.o.f. This may happen at some point of the microscopic wilsonian flow which will correspond also to a point of the space of the emergent effective theory. At this stage the flow towards the IR will be conveniently governed by the universal properties of the new emergent theory.

%%%%%%%%%%%%%%%%%%%%
\subsection{Flow equations of the effective potential}
%%%%%%%%%%%%%%%%%%%%
Let us now consider the basic ingredients of the model of interest here, the reggeon field theory.
In the lowest truncation of the local field approximation, the effective action is  a function 
of the pomeron field $\psi$ and its hermitian conjugate $\psidag$, and it is of the form:
\be
\label{effaction}
\Gamma[\psi^{\dag},\psi]  =\int \!  \, \mathrm{d}^D x \,  \mathrm{d} \tau
 \left( Z( \frac{1}{2} \psidag\partial_{\tau}^{\leftrightarrow} \psi - \alpha' \psidag \nabla^{2} \psi)  + V[\psi^{\dag},\psi] \right),
\ee
where $\tau$ is the rapidity, the dimension $D$ of the transverse space will be mainly specialized to $D=2$, and the potential $V_k$ has the following general properties:\\
(1) $V$ is symmetric under the interchange $\psi \leftrightarrow \psidag$,\\
(2) for real values of $\psi$ and $\psidag$, the real part of $V_k$ is symmetric under $\psi \to -\psi$, $\psidag \to -\psidag$, the imaginary part odd. 
In a polynomial expansion in the region of small fields, this implies that even powers of the field variables come with real coefficients (couplings), 
whereas odd powers have imaginary coefficients. This, in particular, ensures the negative sign of closed Pomeron loops.\\
(3) For small fields our potential can be written, up to an overall constant which we shall often neglect, in the form
\be
\label{potential-factorization} 
V[\psi^{\dag},\psi]=\psidag U(\psidag,\psi) \psi.
\ee
This implies that 
\be
\label{zero-on-axis}
V[0,\psi]=0\,\, \text{and} \,\,  V[\psi^{\dag},0]=0.
\ee
This simple truncation of the effective average action, if we set $Z=1$ and $\alpha'=1$ is called local potential approximation (LPA),
while keeping running $Z$ and $\alpha'$ is known as LPA'.
In the following we will assume that (\ref{zero-on-axis}) will be valid also outside the 
small field approximation.  We shall see that this form is also compatible with the leading asymptotic behavior dictated by the fixed point equation.
To be definite,  for small fields the polynomial expansion has the form: 
\ba
\label{potential}
V[\psi^{\dag},\psi]  =
- \mu \psi^{\dag} \psi + i\lambda  \psi^{\dag}(\psi^{\dag}+\psi)\psi +g  (\psidag \psi)^{2} +g' \psidag ({\psidag}^2 + \psi^2) \psi \nonumber\\
+ i\lambda_5 {\psidag}^2 \left(\psidag + \psi \right) \psi^2 +i\lambda'_5 \psidag \left({\psidag}^3+ \psi^3 \right) \psi +... \,.
\ea  
From this one can see that the RFT has a not Hermitian Hamiltonian even if it can be considered PT symmetric and therefore the Hamiltonian has a real spectrum~\cite{BV}.
Choosing the Pomeron trajectory function  
\be
\alpha(t=-q^2) = \alpha(0) - \alpha' q^2,
\ee
we see the relation between the 'mass' parameter  $\mu$ and the intercept:
\be
\label{intercept}
Z^{-1} \mu=\alpha(0)-1.
\ee
 
We stress that the potential function $V$ should be thought of as a general function 
of the fields $\psi$, $\psidag$. It is convenient to make use of polynomial expansions, but 
since their radius of convergence may be limited one has to go beyond these approximations.

Using
\begin{equation}
 \psi(\tau,\textbf{x})  =\int \!  \, \frac{\mathrm{d}^2 q}{(2\pi)^{D/2}} \,  \frac{\mathrm{d} \omega}{{(2\pi)^{1/2}}} e^{i (\omega \tau-\textbf{x}\cdot \textbf{q})}\tildepsi(\omega,\textbf{q})
\end{equation}
and
\begin{equation}
 \psi^{\dag}(\tau,\textbf{x})  =\int \!  \, \frac{\mathrm{d}^2 q}{(2\pi)^{D/2}} \,  \frac{\mathrm{d} \omega}{{(2\pi)^{1/2}}} e^{-i (\omega \tau-\textbf{x}\cdot \textbf{q})}\tildepsidag(\omega,\textbf{q}),
\end{equation}
we write the kinetic part of the action:
\ba
\Gamma[\psi^{\dag},\psi]_{kin} & =&\frac{1}{2} \int \!  \, \mathrm{d}^2 x \,  \mathrm{d} \tau
 \left( Z( \frac{1}{2} \psidag\partial_{\tau}^{\leftrightarrow} \psi - \alpha' \psidag \nabla^{2} \psi)\right) \nonumber\\
&=& \int d^2q \int d\omega
\Phi(-\omega, q)^{T} G_0^{-1}(\omega,q) \Phi(\omega,q),
\ea
where we have introduced the two-component vectors:
\be
\Phi(\omega,q) = \left( \begin{array}{c} \tildepsi(\omega,q) \\ \tildepsidag(-\omega,q) \end{array} \right)
\ee
with the free propagator matrix:
\ba
\label{propagatormatrix}
G_0(\omega,\textbf{q}) =& \left( \begin{array}{cc}
0 & -i Z \omega + Z \alpha' \bq^{2}\\
iZ \omega + Z \alpha' \bq^{2} & 0
\end{array} \right)^{-1} =&  \left( \begin{array}{cc}
0 &( iZ \omega + Z \alpha' \bq^{2})^{-1}\\
(-i Z \omega + Z \alpha' \bq^{2})^{-1} & 0
\end{array} \right).\nonumber\\ 
\ea

Next let us consider the evolution equation for the effective average action~\footnote{The reggeon field theory model 
is directly mapped to an Hamiltonian model so that one may also use the techniques developed in~\cite{Vacca:2012vt} }. 
We introduce a regulator function $R_k(\omega,q)$ with a cutoff parameter $k$, and
introduce the subscript $k$ to denote the dependence of all quantities upon this regulator. 
This regulator may cutoff the momentum $q$, the energy $\omega$, or both.
Equivalently, we also refer to the 'evolution time' $t=\ln k/k_{0}$: in this notation, the infrared limit is reached for $t \to -\infty$. 
In our matrix notation we introduce :
\be
\mathbb{R}_{k}= \left(
  \begin{array}{cc}
    0 & R_{k} \\
    R_{k} & 0 \\
  \end{array}
\right)
\end{equation}
and
\begin{equation}
\mathbb{\dot{R}}_{k}= \left(
  \begin{array}{cc}
    0 & \dot{R}_{k} \\
    \dot{R}_{k} & 0 \\
  \end{array}
\right),
\ee
where $t=\ln k/k_{0}$ and $\dot{\mathbb{R}}=\partial_{t}\mathbb{R}$ .
The  $k$-dependent effective average action now reads:
\begin{equation}
\Gamma_{k}[\psi^{\dag},\psi]  =\int \!  \, \mathrm{d}^2 x \,  \mathrm{d} \tau
 \left( Z_{k }(\frac{1}{2}\psi^{\dag}\partial_{\tau}^{\leftrightarrow} \psi -\alpha'_{k } \psidag\nabla^{2}\psi) 
+ V_k[\psi, \psidag] \right).
 \end{equation}
%In momentum space the free propagator matrix becomes:
%\be
%G_k(\omega,\textbf{q}) = \left( \begin{array}{cc}
%0 & -i Z_k \omega + Z_k \alpha'_k \bq^{2}+R_k(q)\\
%i Z_k \omega + Z_k \alpha'_k \bq^{2}+R_k(q) & 0
%\end{array} \right)^{-1} .
%\ee

After these definitions we can write down the exact functional RG equation for $\Gamma_{k}$:
\begin{equation}
\partial_{t}\Gamma_{k}=\frac{1}{2}Tr[\Gamma^{(2)}_{k}+\mathbb{R}_k]^{-1}
\partial_{t}\mathbb{R}_k.
\label{eq:exactflow1}
\end{equation}
For constant fields the propagator on the rhs of the flow equation (\ref{eq:exactflow1}) is derived from:
\ba
\Gamma_k^{(2)} +\mathbb{R}_k =  \left( \begin{array}{cc} V_{k \psi \psi} & -i Z_k \omega  + Z_k \alpha'_k q^2 +R_k + V_{k \psi \psidag} \\ i Z_k \omega +    Z_k \alpha'_k q^2 +R_k + V_{k \psidag \psi} & V_{k \psidag \psidag}
\end{array} \right) . 
\label{2pointfunction}
\ea
Inserting the inverse of this into the rhs of (\ref{eq:exactflow1}) we arrive at a partial differential equation for the potential $V_k$ which provides the starting point of our analysis. 

Before turning to results we still need a few further ingredients. First, we define the anomalous dimensions: 
\ba
\label{anomalous-dimensions}
\eta_k=-\frac{1}{Z} \partial_t Z\nonumber \\
\zeta_k= -\frac{1}{\alpha'} \partial_t \alpha'.
\ea
In the following we will also use
\be
z_k=2-\zeta_k.
\ee

Furthermore, we perform a dimensional analysis\footnote{The appearance of dimensions in reggeon field theory should not be confused with physical dimensions. 
In particular, rapidity (which in physical space has no dimensions) plays the role of (dimensionful)  'time' in reggeon field theory.}. We need to distinguish between space and time dimensions:
\be
[x]=k^{-1}
\ee
and
\be
[\tau]=E^{-1}.
\ee
Since the action integral has dimensions zero, we are lead to
\be
[\psi] = [\psidag] = k^{D/2}, \hspace{0.5cm} [\alpha']= E k^{-2}.
\ee
It will be convenient to use dimensionless fields and potential:
\ba
\tilde{\psi} &=& Z_{k}^{\frac{1}{2}} k^{-D/2}\psi \nonumber\\
\tilde{V_k}&=& \frac{V_k}{\alpha'_k k^{D+2}}, \,\,
\tilde{V}_{k \tilde{\psi} \tilde{\psi}} = \frac{\delta^2 \tilde{V_k}}{\delta \tilde{\psi} \delta \tilde{\psi}} = \frac{V_{k\psi \psi}}{Z_k \alpha_k' k^2} .
\label{tildevariables}
\ea
In the polynomial expansion (\ref{potential}) this leads to the definition of dimensionless parameters, e.g.  
\ba
\tilde{\mu}_{k}&=& \frac{\mu_{k}}{Z_{k} \alpha'_{k} k^2} \nonumber\\
\tilde{\lambda}_{k}&=& \frac{\lambda_{k}}{Z_{k}^{\frac{3}{2}}\alpha'_{k}k^2} k^{D/2}\nonumber\\  
\tilde{g}_{k}&=& \frac{g_{k}}{Z_{k}^{2}\alpha'_{k} k^2} k^{D}      \nonumber\\
\tilde{g'}_{k}&=& \frac{g'_{k}} {Z_{k}^{2}\alpha' _{k} k^2 }k^{D}. \nonumber\\
\ea
Finally, we specify our regulator. Clearly there is freedom in choosing a regulator; 
general requiremenst have been discussed, e.g. in \cite{Gies:2006wv}. 
In this first study we make the simple choice of the optimized flat regulator~\cite{Litim},
leaving other regulator schemes 
for future investigations: 
\be
 R_{k}(q)=Z_{k} \alpha'_{k} (k^{2}-q^{2}) \Theta(k^{2}-q^{2})
 \label{eq:regulatorA}
 \ee
 with
  \ba
\dot{R}= \partial_{t} R_{k}(q)&=& -Z_k \alpha'_k (k^2 - q^2) \Theta(k^2 - q^2) [\eta_k+\zeta_k] +  2 k^{2} Z_{k} \alpha'_{k}  \Theta(k^{2}-q^{2}) \nonumber\\
 &=& 2k^2 Z_k \alpha'_{k} \theta(k^2-q^2)\left( 1-\frac{\eta_k+\zeta_k}{2}(1-\frac{q^2}{k^2})\right) .
 \label{eq:regulatordot}
 \ea
Defining 
\be
h_k(q)= Z_k\alpha'_k (q^2 + R_k) = Z_k\alpha'_k(\theta(k^2-q^2) k^2 + \theta(q^2-k^2) q^2)
\label{h-regulator}
\ee
and
\ba
\label{propagator}
G_k(\omega,q)= \left(\Gamma_k^{(2)} +\mathbb{R}\right)^{-1} =\hspace{4cm}\\ 
\frac{1}{V_{k\psi\psi} V_{k\psidag \psidag} - \left( Z_k^2 \omega^2 +(h_k+ V_{k\psi \psidag})^2\right)}
\left( \begin{array}{cc} V_{k \psidag \psidag} & i Z_k \omega  -h_k - V_{k \psi \psidag} \\ -i Z_k \omega  - h_k - V_{k \psidag\psi} & V_{k \psi \psi}
\end{array} \right) ,\nonumber
\ea
and using of (\ref{eq:regulatordot}) we find
\be
\dot{V}_k= 2 Z _k\alpha'_k k^2 \int \frac{d\omega d^Dq}{(2\pi)^{D+1}}\theta(k^2-q^2)
\frac{\left( Z_k\alpha'_k  k^2 + V_{k\psidag \psi}\right) \left( 1- \frac{\eta_k + \zeta_k}{2}(1-\frac{q^2}{k^2})\right)}{Z_k^2 \omega^2 +(h_k+V_{k\psi \psidag})^2 - V_{k\psi\psi} V_{k\psidag \psidag}}.
\ee
Using
\be
\int \frac{d^Dq}{(2\pi)^D} r^{2p} \theta(k^2 - q^2) = \frac{k^{D+2p}}{D+2p} N_D
\ee
with 
\be
N_D=\frac{2}{\sqrt{4\pi}^D \Gamma(D/2)}
\ee
and doing the $\omega$ integration by closing the contour in the complex plane we arrive at:
\ba
\dot{V}_k &=&N_D  A_D(\eta_k,\zeta_k)\alpha'_k k^{2+D} \frac{ \left( 1 + \frac{V_{k\psidag \psi}}{Z_k\alpha'_k  k^2}\right)} 
{\sqrt{1 + 2 \frac{ V_{k\psi \psidag}}{Z_k\alpha'_k  k^2}+
\frac{V_{k\psi \psidag}^2 -  V_{k\psi\psi} V_{k\psidag \psidag}} {(Z_k\alpha'_k  k^2)^2}}},
\label{eq:Vdot}
\ea
where we have introduced the notation
\be
A_D(\eta_k,\zeta_k) = \frac{1}{D} -\frac{\eta_k+\zeta_k}{D(D+2)}.
\ee 
It s convenient to turn to the dimensionless potential $\tilde{V_k}$ introduced in (\ref{tildevariables}): 
\be
 \dot{V}_k =N_D  A_D(\eta_k,\zeta_k)\alpha'_k k^{2+D}  
 \frac{1+ \tilde{V}_{k\tilde{\psi} \tilde{\psidag}} }
 {\sqrt{1+2 \tilde{V}_{k\tilde{\psi} \tilde{\psidag}} +  \tilde{V}^2_{k\tilde{\psi} \tilde{\psidag}} - \tilde{V}_{k\tilde{\psi} \tilde{\psi}}  \tilde{V}_{k\tilde{\psidag} \tilde{\psidag}} }}.
\ee
We use the identity:
\ba
  \frac{\partial V_{k}[\tilde{\psi}]}{\partial t}|_{\tilde{\psi}} &=&  \frac{\partial V_{k}}{\partial t}|_{\psi} +  
 \frac{\partial V_{k}}{\partial\psi}|_{t} \frac{\partial\psi}{\partial t}|_{\tilde{\psi}}+  
 \frac{\partial V_{k}}{\partial\psi^{\dag} }|_{t} \frac{\partial\psi^{\dag} }{\partial t}|_{\tilde{\psi}^{\dag} }\nonumber\\
&=&   \dot{V}_{k}  +  (D/2+\eta/2) ( \tilde{\psi} \frac{\partial V_{k}}{\partial\tilde{\psi}}|_{t}+ \tilde{\psi}^{\dag} \frac{\partial V_{k}}{\tilde{\psi}^{\dag}}|_{t})
\ea
and obtain the following evolution equations:
\be
\dot{\tilde{V_k}}[\tilde{\psi}^{\dag},\tilde{\psi}]=  (-(D+2)+\zeta) \tilde{V_k}[\tilde{\psi}^{\dag},\tilde{\psi}]+(D/2+\eta/2)( \tilde{\psi} \frac{\partial \tilde{V_k}}{\partial \tilde{\psi}}|_{t}+
\tilde{\psi}^{\dag} \frac{\partial \tilde{V_k}}{\partial \tilde{\psi}^{\dag}}|_{t})+ \frac{\dot{V}_{k}} {\alpha' k^{D+2}}.
\label{eq:Vtildedot}
\ee
It should be stressed that this equation is expressed in terms of the dimensionless 
potential, which is a functional of dimensionless fields; for the last term on the rhs this is easily seen from (\ref{eq:Vdot}).
      
Eqs.(\ref{eq:Vdot}) and (\ref{eq:Vtildedot}) are the partial differential  equations for the potentials $V_k$ and $\tilde{V_k}$, resp. They define, for this regulator scheme, 
the basis of our analysis. 

%%%%%%%%%%%%%%%%%%%%%%%%%%%%%%%%%%%%
\section{Search for fixed points (1): zero anomalous dimensions (LPA)}
%%%%%%%%%%%%%%%%%%%%%%%%%%%%%%%%%%%%

The main goal of this paper is the search for fixed points, i.e for solutions $V(\psi,\psidag)$
for which the rhs of (\ref{eq:Vtildedot}) vanishes:
\be
0= (-(D+2)+\zeta) \tilde{V_k}[\tilde{\psi}^{\dag},\tilde{\psi}]+(D/2+\eta/2)( \tilde{\psi} \frac{\partial \tilde{V_k}}{\partial \tilde{\psi}}|_{t}+
\tilde{\psi}^{\dag} \frac{\partial \tilde{V_k}}{\partial \tilde{\psi}^{\dag}}|_{t})+ \frac{\dot{V}_{k}} {\alpha' k^{D+2}}.
\label{fp-equation}
\ee
In the following we will refer to the term proportional $\dot{V}$ as the 'quantum' part, 
to the terms in front as the 'canonical' part. 
This nonlinear partial differential equation 
can be solved only approximately, and in order to obtain a consistent picture we have to make use of several different methods.
Throughout this paper we restrict ourselves to constant fields $\psi$ and $\psidag$. This is consistent with our truncation.
As a general strategy, we first will set the anomalous dimensions equal to zero.
In a second step we will generalize to nonvanishing anomalous dimensions; this requires the  approximate calculation of anomalous dimensions
(from the 2-point function) which will be described in the first part of the the following section. 
We will find that the presence of the anomalous dimensions does not alter the qualitative shape of the fixed point potential obtained from the case without 
anomalous dimensions; on the other hand, nonzero anomalous dimensions tend to make calculations technically more complicated. 
For the remainder of this section we will set the anomalous dimensions $\eta$ and $\zeta$ equal to zero.
In the following section we allow for nonvanishing anomalous dimensions.                

In this and in the following section all our analysis we will done for dimensionless variables, i.e. we consider the dimensionless potential depending on dimensionless fields.   
For simplicity from now on we will drop the tilde-notation which we have introduced in order to distinguish between 
dimensionful and dimensionless variables. 
Only in section 5 we will distinguish between the two sets and re-introduce the tilde-symbols.     

%%%%%%%%%%%%%%%%%%%%
\subsection{Approximation schemes}
%%%%%%%%%%%%%%%%%%%%
Let us briefly outline various approximation schemes. First, we perform polynomial expansions of the potential function $V$. 
The most obvious point to expand around is the origin, i.e. the point $(\psi,\psidag)=(0,0)$ (which is stationary, i.e. solution of the equations of motion for constant fields).
This expansion is written in ( \ref{potential}). 
Inserting this ansatz on both sides of (\ref{eq:Vtildedot}) and equating equal powers of the fields,
one obtains differential equations for the coupling constants $\mu$, $\lambda$, etc. which is a projection of the non linear differential equation on the basis of monomials in the fields, which further affects the approximation chosen for the theory space, which can be refined with increasing the order.
The rhs of these equations define the coupled set of $\beta$-functions, and their zeroes define fixed points of the flow equations. As usual, one proceeds with truncations:
the lowest truncation has only the two parameters $\mu$ and $\lambda$, in the next truncation one includes the quartic couplings  $g$ and $g'$ and so forth. As an example, in Appendix A we have listed the $\beta$-function for the quartic truncation.
An important step is the stability analysis of fixed points: we have to investigate the matrix of derivatives of the $\beta$-functions
and compute the eigenvalues and eigenvectors.  
A positive (negative) eigenvalue is repulsive (attractive) for $t \to \infty$; since we are interested in the infrared limit, $t \to -\infty$, the directions are reversed.    

It is often useful to introduce, instead of $\psi$, $\psidag$, other variables which respect, as much as possible, the global symmetry properties  of the potential. 
In our case we may use:
\be
r=\psidag \psi, \,\,u=\psi+\psidag.
\label{r-u-variables}
\ee
With these variables, we can write a completely equivalent expansion around the origin;
\be
V=\sum_{n,m\ge0} \lambda_{n,m}r^{1+n} (i u)^m
\label{potential-r-u}
\ee
with real-valued constants $\lambda_{n,m}$.
The beta functions generated by these two expansions,  (\ref{potential}) and (\ref{potential-r-u}), are linearly related.

For the lowest truncation it has been known since a long time that the effective potential $V$
has several stationary points:
\be
(\psi_0,\psidag_0)= (0,0),\hspace{0.5cm} (\frac{\mu}{i\lambda},0),\hspace{0.5cm} (0, \frac{\mu}{i\lambda}), \hspace{0.5cm}
(\frac{\mu}{3i\lambda},\frac{\mu}{3i\lambda}).
\label{stat-points}
\ee
The last point lies on the diagonal $\psi=\psidag$ and is a minimum  of the potential in the space of fields with imaginary values, whereas  
the other ones lie on the $\psi$ or $\psidag$ axis and represent saddle-points.
All these points lie in the subspace spanned by $\Im\psi$ and $\Im \psidag$ (note that, as a function of
$\Im \psi$ and $\Im \psidag$, the minimum on the diagonal becomes a maximum), more precisely at negative values of $\Im \psi$ and $\Im \psidag$. This suggests (and later on will be confirmed) that this part of the field space plays an essential role.
Let us stress that the pomeron field, from its physical meaning, is known to be mainly imaginary but with small real corrections.

One can easily see that such extrema will exist also for higher truncations. It is often the existence of extrema away from the origin which leads to a  slow convergence of the expansion around the origin. Our numerical investigation of the expansion around 
$(\psi,\psidag)=(0,0)$, in  fact,  show that the convergence with increasing order of truncation is slow. It is therefore useful to try different expansions which may lead to a better approximation to the global solution of the fixed point PDE for the potential. Generically we write
\be
V=\sum_{n,m} V_{nm} (\psi-\psi_0)^n (\psidag-\psidag_0)^m. 
\label{exp-on-axis}
\ee  
When deriving the $\beta$-functions for this ansatz, we first observe that the new
parameters $\psi_0$ and $\psidag_0$ replace two  of the other couplings. Furthermore, the requirement that $(\psi,\psidag)=(\psi_0,\psidag_0)$ is a stationary point of $V$, leads to a modification of the $\beta$-functions for $\psi_0$ and $\psidag_0$:
\be
0=\left( \begin{array} {c} \dot{V}_{\psi}\\\dot{V}_{\psidag} \end{array}  \right)
+\left( \begin{array} {cc} V_{\psi\psi} & V_{\psidag\psi}\\
V_{\psi\psidag}&V_{\psidag\psidag} \end{array}  \right)
 \left( \begin{array} {c} \dot{\psi_0}\\\dot{\psidag_0} \end{array}  \right).
\label{comoving}
\ee
This equation expresses the condition that the stationary point of the potential is comoving with the flow.     
Guided by (\ref{stat-points}), we consider an expansion around a configuration on the $\Im\psi$-axis which may be a non trivial stationary point. In terms of the symmetric variables (\ref{r-u-variables})
such a point has the coordinates $r_0=0$ and $u=u_0$, and the polynomial expansion has the form:
\be
V=\sum_{n+m>0} \lambda_{n,m}r^{1+n} (i (u-u_0))^m.
\label{Vaxis}
\ee
The new parameter $u_0$ replaces one of the other couplings, preferably $\mu=-\lambda_{00}$,
and (\ref{comoving}) leads to 
\be
\dot{u}_0= - \frac{\dot{V}_r}{V_{ru}}.
\ee
As expected, our numerical analysis will show that this expansion around an extremum away from the origin has better convergence properties than the expansion around the point of zero fields.

As a third tool of analyzing the fixpoint equations, we search for numerical solutions of differential equations. 
For this part of our analysis we restrict ourselves to the imaginary parts of our field variables.  We put 
 $\psi=i\phi$, $\psidag=i\phidag$ and define the combinations:  
\be
\phi_{\pm} = \phi \pm \phidag.
\label{variables-pm}
\ee
 Rather than trying to solve the partial differential equations 
(\ref{fp-equation}) we consider truncations along the $\phi$-axis:
\be
V=\phi \phidag  f_{a1}(\phi_+) + (\phi \phidag)^2 f_{a2}(\phi_+) + (\phi \phidag)^3 f_{a3}(\phi_+)+...\,.
\label{diff-axis}
\ee
By symmetry, the same ansatz applies also to the $\phidag$-axis.   
Along the diagonal $\phi=\phidag$ we make the ansatz
\be
V=f_{d0}(\phi_+) +\phi_{-}^2 f_{d2}(\phi_{+}) +\phi_{-}^4 f_{d4}(\phi_{+} )+...\,.
\label{diff-diag}
\ee 
(because of the symmtry of V under $\psi \leftrightarrow \psidag$ we allow only for even 
powers in $\phi_-$). Inserting the ansatz (\ref{diff-axis}) into  (\ref{fp-equation}) we derive a coupled set of second order differential equations for $f_{a1},f_{a2},...$ which can be solved numerically. The same applies to the ansatz (\ref{diff-diag}).

For this analysis it is important to specify the behavior of the solutions at the origin and at infinity. For small fields the form of the potential is given by the expansions  
(\ref{potential}) or (\ref{potential-r-u}).  For asymptotically large fields we observe in (\ref{fp-equation}) (with zero anomalous dimensions) that
\be
V(\psi,\psidag) \sim \left( \psi \psidag \right)^2 V_{\infty}(\psi/\psidag) \left(1+ {\cal{O}} (1/u^4)\right)
\label{large-field-behavior}
\ee
provides a solution where $V_{\infty}(\psi/\psidag)$ is some unknown function which only depends upon the ratio of the two fields. This behavior is the canonical one, i.e. in (\ref{fp-equation}) it solves the first 'canonical' part, whereas the 'quantum' part (proportional to 
$\dot{V}$) provides subleading corrections. The behavior (\ref{large-field-behavior}),
when inserted into (\ref{diff-axis}) or  (\ref{diff-diag}) determines the asymptotic behavior 
of the functions $f_i()$. A closer look at (\ref{fp-equation}) shows that another 
'noncanonical' asymptotic behavior is possible:
%\be
%V(\psi,\psidag) =\phi \phidag \left(1+ {\cal O} (1/\phi_+)\right) =-\psi \psidag \left(1+ {\cal O} (1/u)\right) .
%\label{large-field-special}
%\ee
\be
V(\psi,\psidag) =-\psi \psidag \left(1+\frac{a_1}{u}\left(1+\frac{\psi \psidag}{u^2}+\frac{1}{(8\pi)^2 u^4}\right)+\cdots \right) =\phi \phidag \left(1+ {\cal O} (1/\phi_+)\right) .
\label{large-field-special}
\ee

In this case we observe in the differential equation, for large field values, a rather subtle cancellation between the canonical part and the quantum part.    

%%%%%%%%%%%%%%%%%%%%%%
\subsection{Numerical results}
%%%%%%%%%%%%%%%%%%%%%%

Let us now report results of our numerical analysis. We begin with the polynomial around 
the origin. Denoting, on the rhs of (\ref{fp-equation}), the coeficients of the expansion by
$\beta_{\mu}(\mu,\lambda, g,g',...)$, $\beta_{\lambda}(\mu, \lambda,g,g'...)$ and defining the vector $\bbeta$:
\be
\bbeta=(\beta_{\mu}, \beta_{\lambda},...),
\ee
we search for zeroes of $\bbeta$, the fixed points in the space of the couplings. For the stability analysis we compute eigenvalues and eigenvectors of the matrix of derivatives
\be
 \Big[\frac{\partial \bbeta}{\partial \mu},\frac{\partial \bbeta}{\partial \lambda},...\Big]|_{\text{fixed point}}.
\ee

First, there is a fixed point at the point of zero coupling, which, in the infrared limit,  is repulsive. 
Next, as to be expected, there are many fixed point solutions which are not robust when changing the truncation, i.e. they come and go if we turn to higher and higher truncations.
In general, these fixed points have positive, negative and even complex-valued eigenvalues. Finally and most important, we find, for all truncation one fixed 
point which is robust and always has the same stability properties: one relevant direction,
i.e. one negative eigenvalue with all the other ones being real and positive. Let us list a few features of this fixed point, in particular as a function of the different truncations. In Table 1,
'truncation 3' denotes the polynomial expansion which retains quadratic and cubic terms ($\mu$ and $\lambda$); 'truncation 4' includes also quartic terms, and so forth. 
In the first row of Table 1 we list, for different truncations,  the critical exponent  $\nu$ which is defined as the negative inverse of the negative eigenvalue:\\
\begin{center}
\begin{tabular}{|l|c|c|c|c|c|c|c|c|c|c|}\hline
\text{truncation} &3&4&5&6&7&8&9&10&11&12\\ \hline
\text{exponent} $\nu$ & 0.53&0.59&0.59&0.78&0.76&0.72&0.72&0.74&0.74&0.73\\ \hline
\text{mass $\tilde{\mu}$} &0.111&0.274&0.386&0.429&0.341&0.388&0.388&0.400&0.399&0.397
\\ \hline
\text{ $i\psi_{0,diag}$}&0.035&0.059&0.072&0.078&0.075&0.073&0.073&0.074&0.074&0.074
\\ \hline
\text{ $i\psi_{0,axis}$}&0.106&0.175&0.214&0.228&0.221&0.215&0.217&0.219&0.219&0.218
\\ \hline
\end{tabular} \\ \vspace{0.5cm}
Table 1: Polynomial expansion around $(\psi,\psidag)=(0,0)$.\\
Parameters of the fixed point for different truncations.
\end{center}     
For comparison, the critical value obtained from a Monte Carlo simulation~\cite{MC1,MC2}
 is $\nu=0.73$~\footnote{It has been shown long ago by J. Cardy~\cite{Cardy:1980rr}
that reggeon field theory belongs to the same universality class as the non equilibrium critical phenomena known as directed percolation.
This is a process which has been intensively studied and has been investigated more recently also with functional renormalization group techniques~\cite{Canet:2003yu}.}.
For a bit more extended comparison with Monte Carlo results, also for the case with anomalous dimensions taken into account, see Appendix D.
In the second row we give the fixed point values for the 'mass' $\mu$. The convergence is rather slow: below the order 7 we observe rather strong changes. We also look for stationary 
points, i.e. points in field space where both $\frac{\partial V}{\partial \psi}$ and 
$\frac{\partial V}{\partial \psidag}$ vanish. 
In each truncation, we find several such stationary points. The ones closest to the 
origin have the same characteristics as the ones listed  (\ref{stat-points}). 
In the third row we present the numerical values $\psi_{0, diag}=\psidag_{0,diag}$ of the nearest stationary point on the diagonal, in the fourth row the value $\psi_{0,axis}$ of the nearest stationary point on the $\psi$ axis. 

Next we study the polynomial expansion around a stationary point on the $\psi$ axis, as written in (\ref{exp-on-axis}). 
The $\beta$-function for the parameter $u_0$ has been described above. For the stability analysis of the fixed point it is important to note that,
on the rhs of  (\ref{eq:Vtildedot}), the expansion in powers of the field variables 
encounters time derivatives of the couplings and of the parameter $u_0$. 
This leads to slightly modified equations of motion for the couplings. Our findings for fixed points are similar to those of the polynomial expansion around  the origin; for all truncation we find 
the fixed point with one relevant direction which turns out to be robust with respect to 
varyjng the truncation.  For this expansion it is convenient to define the 
effective reggeon mass and triple coupling:
\ba
\mu_{eff}& =& -\frac{\partial^2 V}{\partial \psi \partial \psidag}|_{\psi=\psidag=0}= 
                   -\frac{\partial V}{\partial r}|_{r=u=0}
\nonumber\\
\lambda_{eff}&=& -i \frac{\partial^2 V}{\partial r \partial u}|_{r=u=0}.
\label{effective}
\ea  
Numerical results are shown in Table 2.
\\
\begin{center}
\begin{tabular}{|l|c|c|c|c|c|c|}\hline
\text{truncation} &3&4&5&6&7&8\\ \hline
\text{exponent} $\nu$ & 0.74&0.75&0.73&0.73&0.73&0.73\\ \hline
\text{mass $\tilde{\mu}_{eff}$} &0.33&0.362&0.384&0.383&0.397&0.397
\\ \hline
\text{ $i\psi_{0,diag}$}&0.058&0.072&0.074&0.074&0.0.074&0.074
\\ \hline
\text{ $iu_0$}&0.173&0.213&0.218&0.218&0.218&0.218
\\ \hline
\end{tabular} \\ \vspace{0.5cm}
Table 2: Polynomial expansion around $(r,u)=(0,u_0)$.\\
Parameters of the fixed point for different truncations.
\end{center}   
For this fixed point solution we also compute the stationary point on the diagonal which is 
closest to the origin (third line).   
A comparison with the expansion around the origin shows that now the 
series of truncations converge faster. In addition, for the expansion around the origin 
the domain of convergence contains the nearest extrema on the $\psi$-axis and on the diagonal,
and the expansion around $u_0$ includes the origin. This implies, in particular,
that the expansion around $u_0$, which converges faster than the expansion around the 
origin, conveniently gives a reliable description of the potential in the vicinity of zero fields.

In order to illustrate these findings of the polynomial expansions we show, in Fig.\ref{cubic}a,
the flow of couplings for the cubic truncation, as obtained from the expansion around 
the stationary point on the $\psi$ axix. 
Although the numerical values of this low-order truncation are not accurate yet, it nevertheless correctly illustrates the flow in the space of couplings:
the origin is unstable, i.e. in the infrared limit all trajectories leave the fixed point. The other fixed point is our 
candidate: there is the distinguished relevant outgoing direction (in green) and the other incoming direction (in red).
In higher truncations where the space of coupling becomes n-dimensional, the 
outgoing direction remains one-dimensional whereas the space of incoming trajectories  
becomes (n-1)-dimensional. In Fig.~\ref{cubic}b we show a qualitative view of a (fictitious) truncation with 3 couplings:
the space of incoming trajectories becomes a 2-dimensional plane.  In the following we name this surface 'critical subspace'.
 
\begin{figure}                 
\begin{center}
\includegraphics[width=5cm,height=5cm]{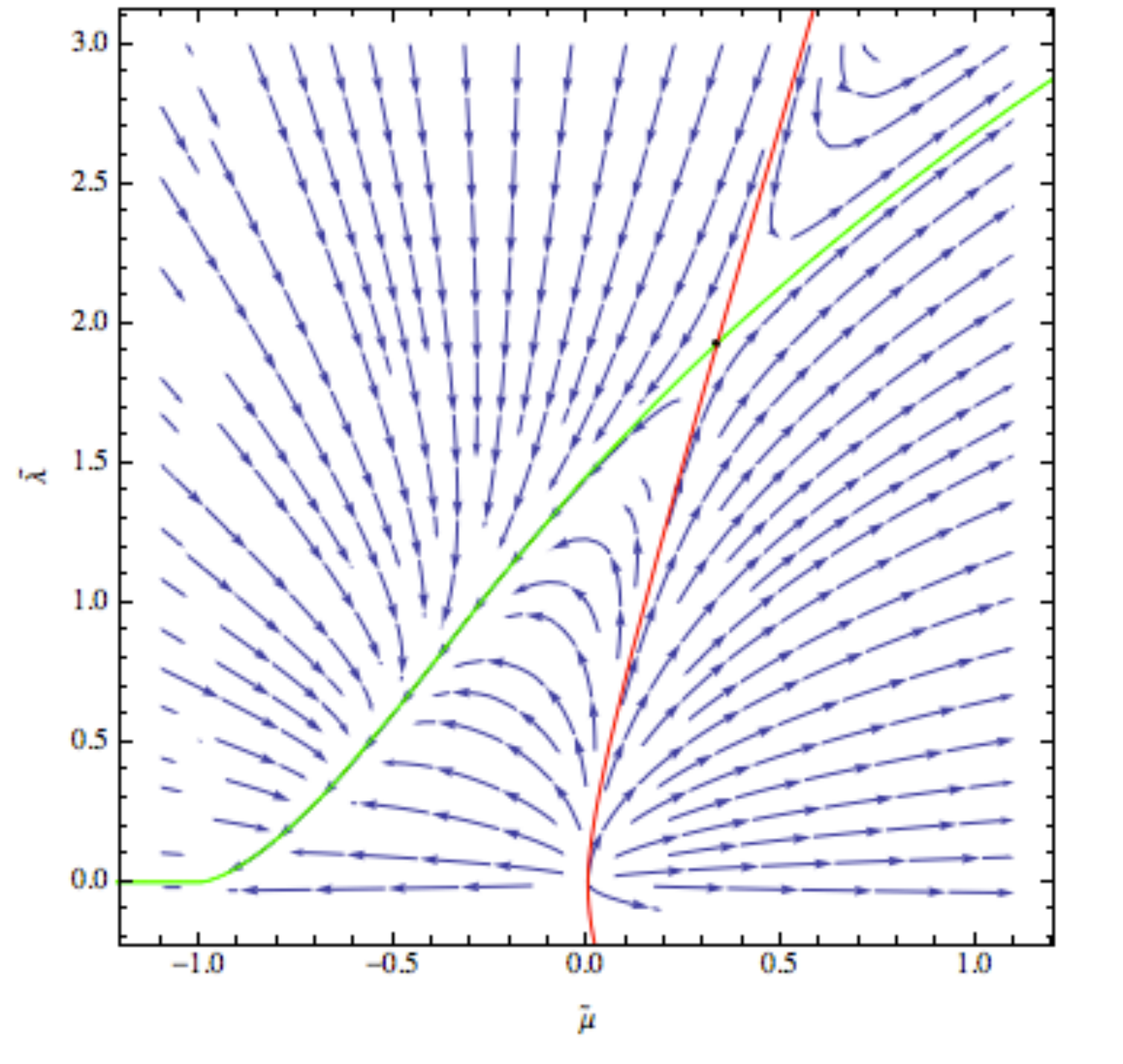}\hspace{2cm}
\includegraphics[width=5cm,height=5cm]{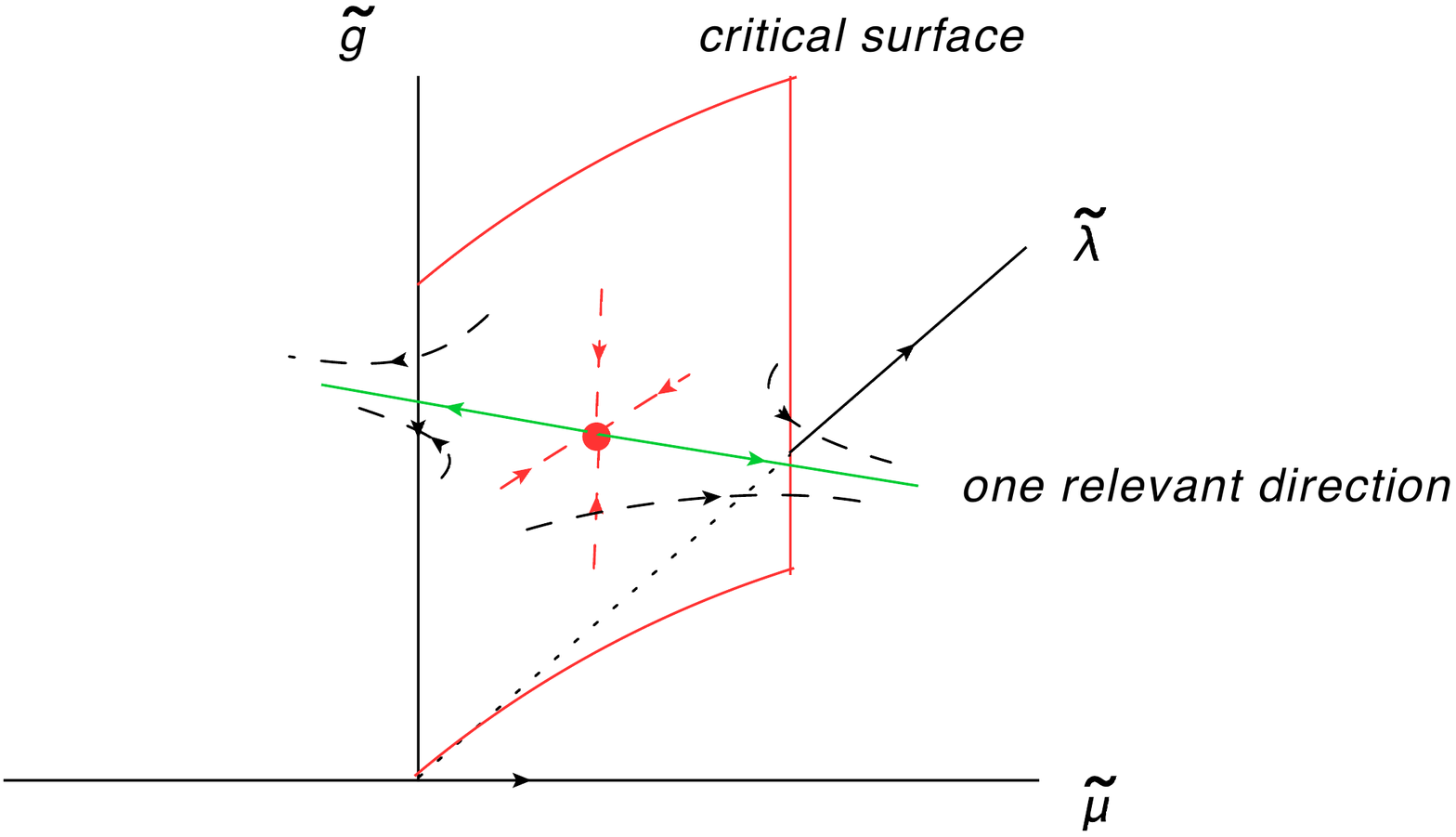}
\caption{(a) Flow in a cubic truncation (without anomalous dimensions), (b) flow in a space with three couplings.}
\label{cubic}
\end{center}
\end{figure}

So far we have concentrated on the shape of the fixed point potential in the region of small fields; in particular we have looked at the stationary points closest to the origin, generalizing the results in (\ref{stat-points}). In order to study the potential for larger fields we make use 
of the differential equations, starting from the ansatz (\ref{diff-axis}) or (\ref{diff-diag}).
In both cases we used a truncation of third order and solve numerically for the functions
$f_{ai}$ and $f_{di}$. Some details are described in the appendix B.

\begin{figure}
\begin{center}
\includegraphics[width=5cm,height=5cm]{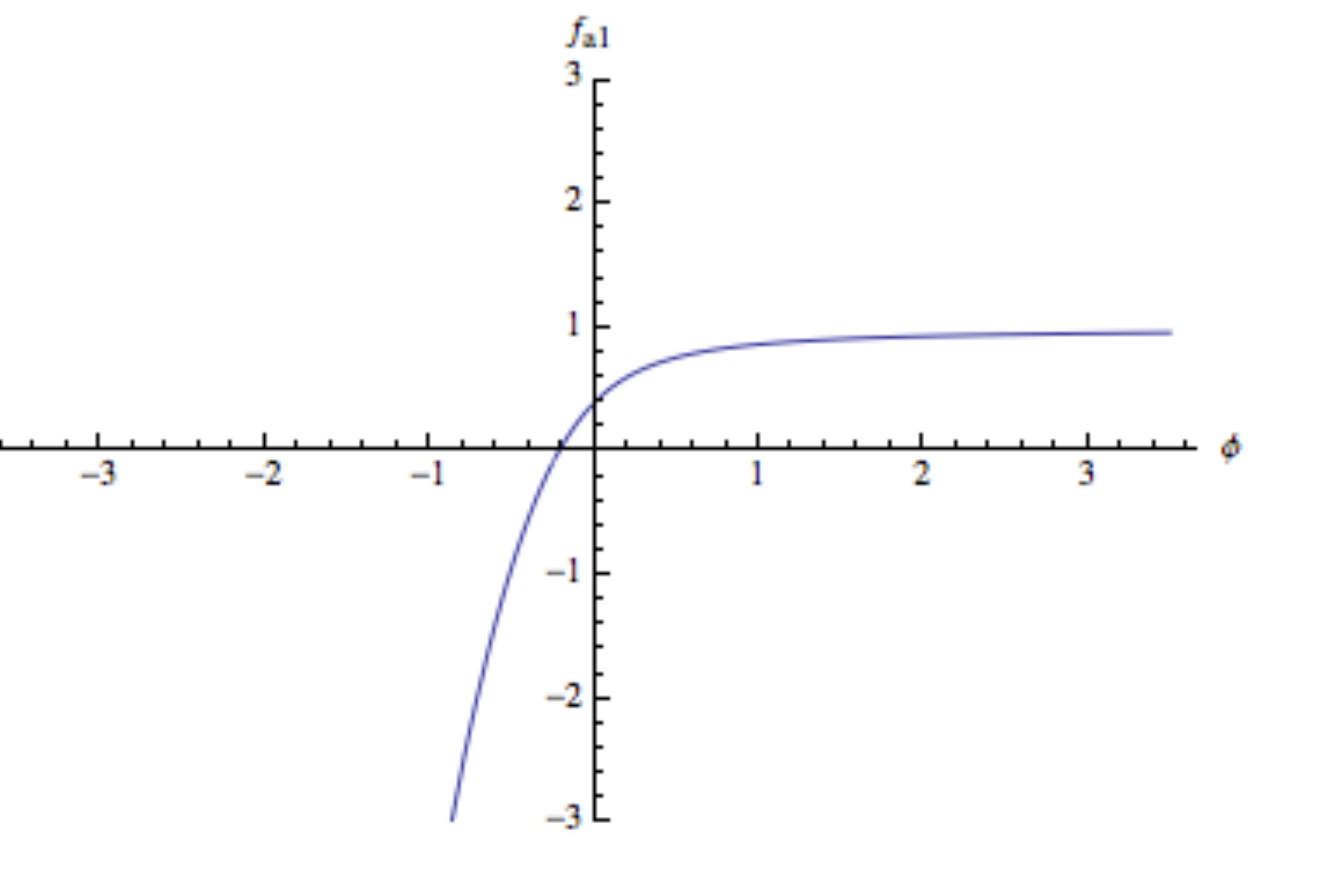}\vspace{3cm}
\includegraphics[width=5cm,height=5cm]{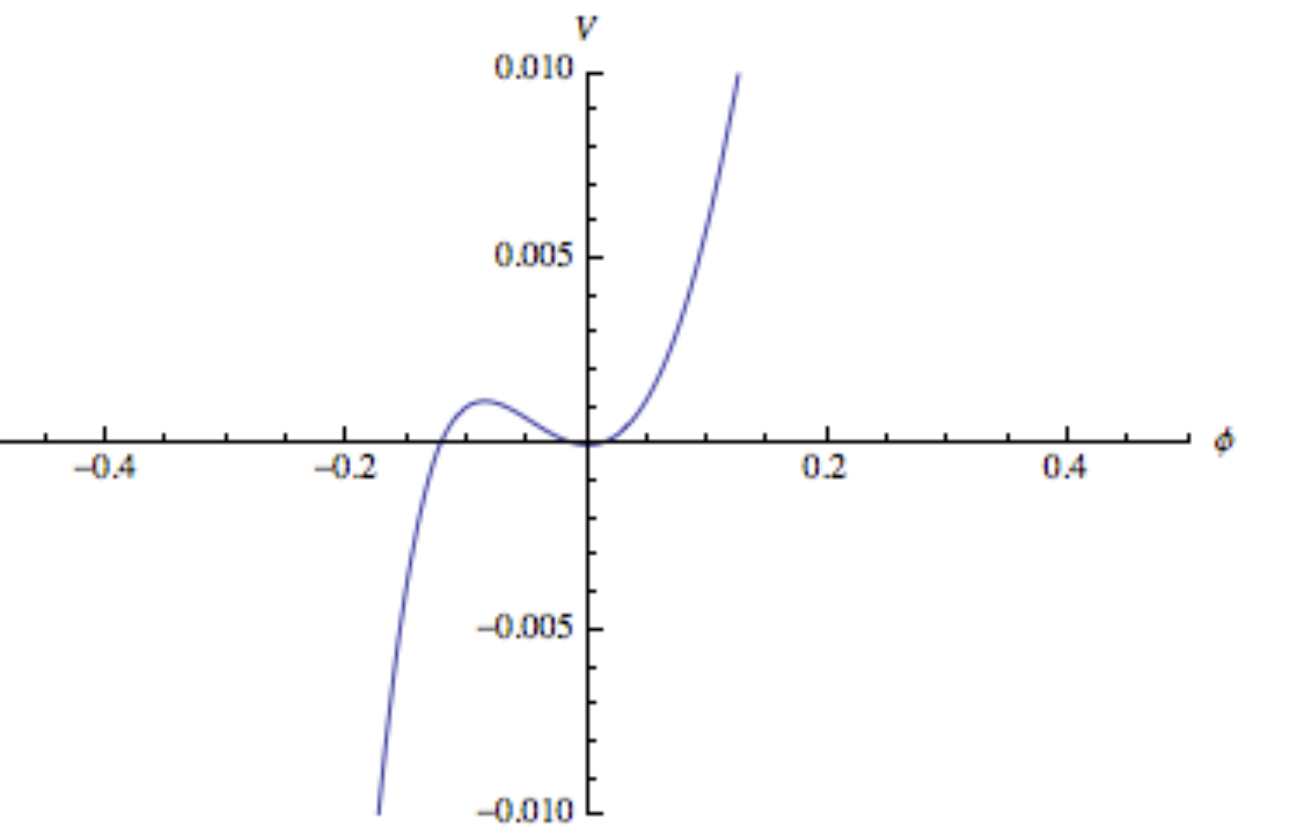}
\caption{ (a) The function $f_{a1}$ on the axis $\phidag=0$; (b) shape of the potential along the diagonal $\phi=\phidag$.}
\label{diff}
\end{center}  
\end{figure}

In the vicinity of the $\phi$ axix  (i.e. small $\phi\phidag$) we show, as a result of our calculation,  
the derivative of $f_{a1}(\phi_+)$ at $\phi_+=0$: a stationary point of the potential function $V$ on the $\phi$ axis must have $\frac{\partial V}{\partial \phi}=\frac{\partial V}{\partial \phidag}=0$. When written in terms of the variables 
$\phi\phidag$ and $\phi_+=\phi+\phidag$ and applied to the ansatz (\ref{diff-axis}), this condition means that $f_{a1}$ must have a zero in  $\phi_+$. Our result for  $f_{a1}(\phi)$
is shown in Fig.~\ref{diff}a: we see the stationary point at $\phi_+=-0.22$ and no other zero, i.e. there is no other stationary point on the axis.

Along the diagonal line $\phi=\phidag$ we use the variables $\phi_+$ and $\phi_-$, with the ansatz 
given in (\ref{diff-diag}). Again some details are described in the Appendix B. As the main result of these calculations, we show in Fig.~\ref{diff}b the shape of the potential along the diagonal. Apart from the maximum at $\phi_+=-0.07$ the potential is a monotonic function without any further structure.        

We summarize this part of our analysis by showing in a 3-dimensional plot the 
shape of the effective potential at the fixed point (Fig.~\ref{pot3D}).  

\begin{figure}
\begin{center}
\includegraphics[width=9cm,height=5cm]{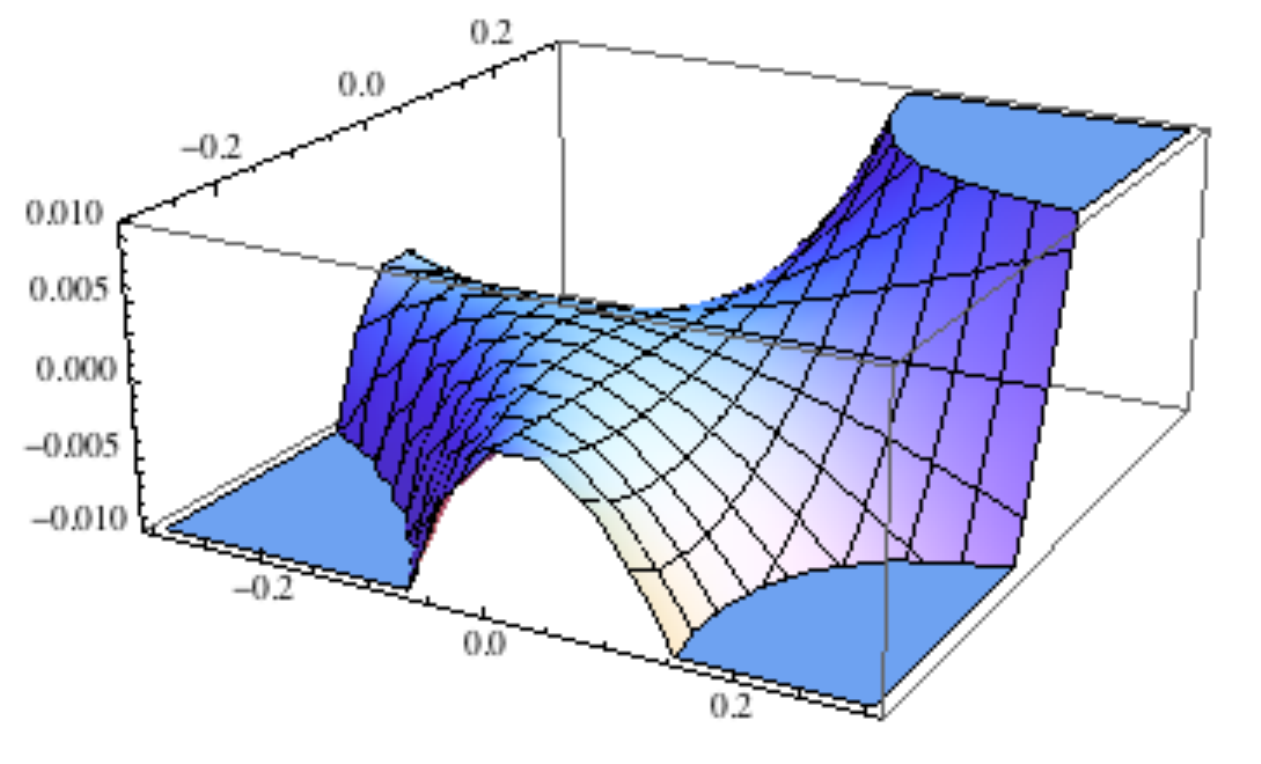}
\caption{Shape of the potential function $V$ as a function of $\phi=\Im \psi$ and $\phidag=\Im \psidag$.}
\label{pot3D}
\end{center}  
\end{figure}

Let us conclude with a brief summary of this section. The main result is the existence 
of a fixed point. In the multi-dimensional space of parameters (couplings) this fixed point 
has one relevant deformation (negative eigenvalue, i.e. repulsive in the infrared limit), all other 
directions have positive eigenvalues (attractive in the infrared limit). The possible physical 
significance of these stability properties will be discussed further below. 
We have investigated the fixed point potential in some detail, and we conclude that the potential has a rather simple shape.
In detail, the shape may depend on our choice of the coarse-graining regulators, and  it may change if we chose another regulator.  The qualitative features, however, should not change.
Apart from the three stationary points close to the origin the potential is monotonically rising (falling) if we restrict the plots to the pure imaginary values for the fields.
The structure of the stationary points is (qualitatively) the same as derived in perturbation theory from the lowest truncation many years ago using the lowest possible truncation in the vicinity of $D=4$ \cite{Amati:1975fg}.

%%%%%%%%%%%%%%%%%%%%%%%%%%%%%%%%%%%%%%%%
\section{Search for fixed points (2): including anomalous dimensions (LPA')}
%%%%%%%%%%%%%%%%%%%%%%%%%%%%%%%%%%%%%%%%

So far we have restricted our analysis to zero anomalous dimensions. In this section 
we compute the anomalous dimensions and include them into our numerical analysis.
We will be guided by the general experience that, as long as the anomalous dimensions 
are not too large, they leave the results obtained on neglecting them qualitatively intact.
As we will see, our numerical results support these expectations.    

%%%%%%%%%%%%%%%%%%%%%%%%%%%
\subsection{Calculation of anomalous dimensions}
%%%%%%%%%%%%%%%%%%%%%%%%%%%

In order to obtain the anomalous dimensions we need to compute vertex functions, i.e. functional derivatives of the effective action. We expand in  powers of the field variables around the field values $\psi_0,\psidag_0$. The  one-particle irreducible n-point vertex functions  are defined as
\be
 \Gamma_{k}^{(n, m)}=\frac{ \delta^{n+m}\Gamma_{k}[\psi^{\dag},\psi]}{\delta \psi_{1}^{\dag}(x_1)\ldots \delta \psi_{n}^{\dag}(x_n)\delta \psi_{1}(y_1) \ldots \delta \psi_{m}(y_m)}|_{\psi_0,\psidag_0}.
\ee
For the anomalous dimensions $\eta$ and $\zeta$ it will be enough to consider the two point vertex function $\Gamma^{(1,1)}$ which is obtained by taking two derivatives. In the one loop approximation we have: 
\ba
\partial_{t}\Gamma_{k}^{(1,1)}[1,2]=\frac{1}{2}  Tr \left(  \partial_{t} \mathbb{R}_k  \frac{1}{[\Gamma^{(2)}_{k}+\mathbb{R}_k]}  \Gamma^{(2)}_{k,1} \frac{1}{[\Gamma^{(2)}_{k}+\mathbb{R}_k]}\Gamma^{(2)}_{k,2}  \frac{1}{[\Gamma^{(2)}_{k}+\mathbb{R}_k]}+(1\leftrightarrow 2)\right) \nonumber\\
-\frac{1}{2} Tr \left( \partial_{t} \mathbb{R}_k  \frac{1}{[\Gamma^{(2)}_{k}+\mathbb{R}_k]} \Gamma^{(4)}_{k,12} \frac{1}{[\Gamma^{(2)}_{k}+\mathbb{R}_k]} \right).
\label{floweq}
\ea
Here $\Gamma^{(2)}_{k \psi}$,  $\Gamma^{(2)}_{k \psidag}$, and $\Gamma^{(2)}_{k \psi\psidag }$ denote the 2x2 matrices obtained by
taking first and second derivatives of $\Gamma^{(2)}$ and putting $\psi=\psi_0$,
$\psidag=\psidag_0$ afterwards. The propagator matrix $[\Gamma^{(2)}_{k}+\mathbb{R}_k]^{-1}$ 
was defined in (\ref{propagator}), and we substitute $\psi=\psi_0$,
$\psidag=\psidag_0$.
This  flow equation  can diagrammatically be illustrated in Fig.4:
\begin{center}
\includegraphics[width=5cm,height=5cm]{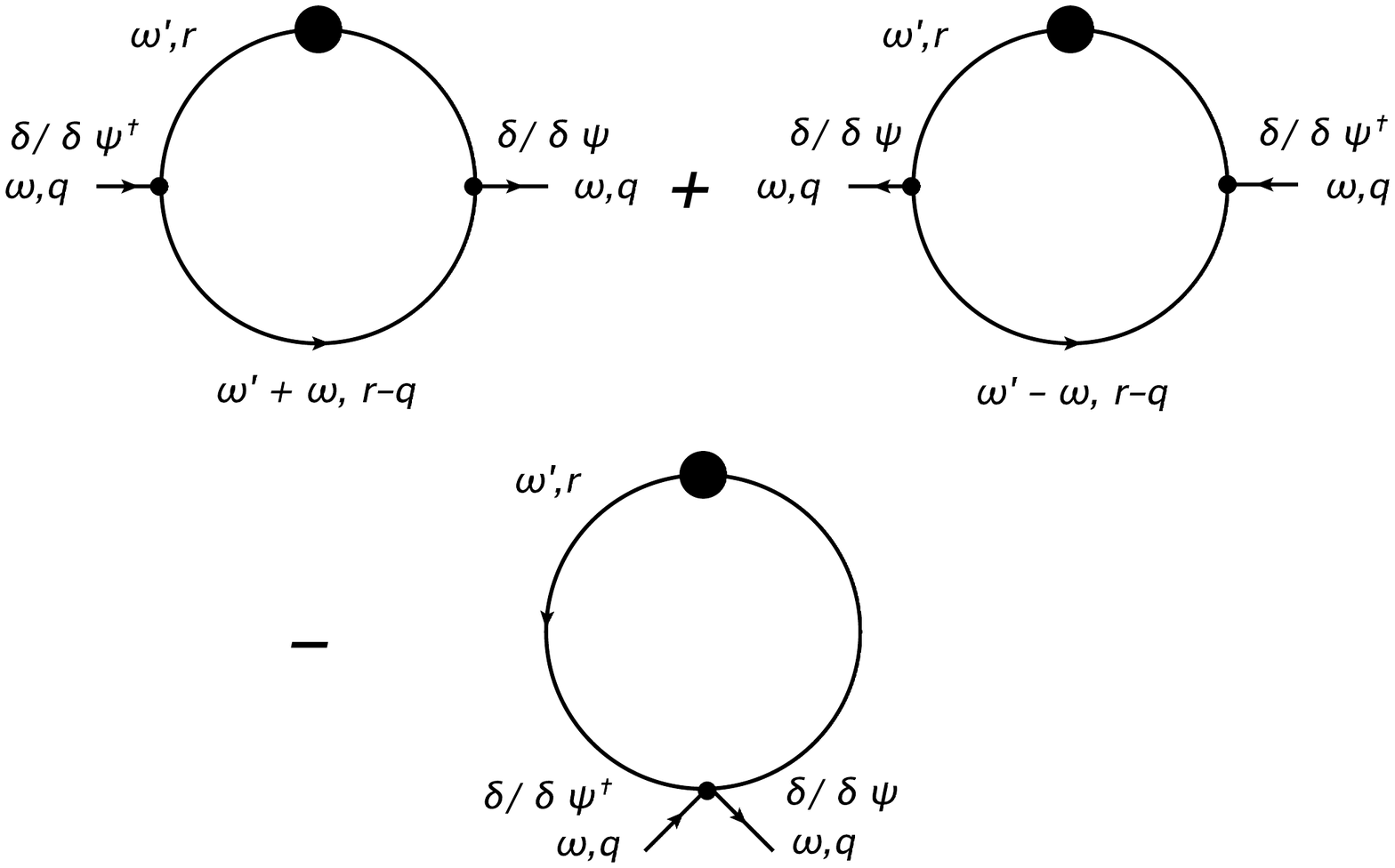}\\
Fig.4: Illustration of the two-pointfunction $\Gamma_{k}^{(1,1)}$: the big blob represents the regulator insertion,  $ \partial_{t}R(k)$,
lines the propagator $ [\Gamma^{(1,1)}_{k}+R(k)]^{-1} $, and small dots the triple or quartic interaction vertices.
\end{center}
On the lhs of the flow equations the functional derivatives lead to:
\be
\Gamma_k^{(1,1)} =\frac{\delta^2}{\delta \psi \delta \psidag} \Gamma_k |_{\psi=\psi_0,\psidag=\psidag_0} =
i Z_k \omega + Z_k \alpha'_k q^2 + \frac{\delta^2 V_k}{\delta \psi \delta \psidag}
 |_{\psi=\psi_0,\psidag=\psidag_0} .
\ee
Denoting the rhs of (\ref{floweq}) by $I^{1,1}(\omega,q)$ we arrive at the differential equation:
\be
\dot{\Gamma}^{(1,1)}_k = I^{1,1}(\omega,q).
\ee
In order to find the evolution equations for the anomalous dimensions we use:
\be
i Z_k=\lim_{\omega\to 0, q\to 0} \frac{\partial}{\partial \omega} \Gamma_k^{(1,1)}(\omega,q)
\ee
and
\be
Z _k \alpha'_k = \lim_{\omega\to 0, q\to 0}  \frac{\partial}{\partial q^2} \Gamma_k^{(1,1)}(\omega,q).
\ee
This leads to
\be
-i \eta_k = \frac{1}{Z_k} \lim_{\omega\to 0, q\to 0}  \frac{\partial}{\partial \omega} I^{(1,1)}(\omega,q)
\ee
and
\be
-\eta_k - \zeta_k = \frac{1}{Z_k \alpha'_k} \lim_{\omega\to 0, q\to 0} \frac{\partial}{\partial q^2} I^{(1,1)}(\omega,q),
\ee
i.e. we need the integral $I^{1,1}(\omega,q)$ and its derivatives with respect to $\omega$ and $q^2$ at the point $\omega=0, q=0$.
Note that, within our truncation, the last term in (\ref{floweq}) is independent of $\omega$ and $q$ and thus does not contribute to these derivatives.

In D-dimensional momentum space the trace on the rhs of  (\ref{floweq}) reads as follows:
\ba
I^{(1,1)}(\omega,q)= \int \!  \, \frac{\mathrm{d}^D q'   \mathrm{d} \omega'} {(2\pi)^{D+1}} \,   Tr \,\Big[\partial_{t}\mathbb{R}_k(q')  
 G_k(\omega',q')  \Gamma^{(2)}_{k \psidag} G_k(\omega+\omega',q+q')  \Gamma^{(2)}_{k \psi}
 G_k(\omega',q') \Big].\nonumber\\
 \label{eq:flowgamma2}
\ea
Here we have made use of the fact that the two terms in the first line on the rhs of 
 (\ref{floweq}) coincide.
The Green's function $G(\omega,r)$ was defined in  (\ref{propagator}),
and we find it convenient to separate the denominator $D$ from the matrix part $N$ at numerator:
\be 
G(\omega,q) = \frac{1}{D(\omega,q)} N(\omega,q),
\ee 
where 
\be
D(\omega,q)= V_{k\psi\psi} V_{k\psidag \psidag} - \left( Z_k^2 \omega^2 +(h_k+ V_{k\psi \psidag})^2\right)
\ee
and
\ba
N(\omega,q)= \left( \begin{array}{cc} V_{k \psidag \psidag} & i Z_k \omega  -h_k - V_{k \psi \psidag} \\ -i Z_k \omega  - h_k - V_{k \psidag\psi} & V_{k \psi \psi}
\end{array} \right).
\ea
Let us begin with the $\omega$-derivative. The $q'$-integral (at $q=0$) is fairly straightforward: 
because of the $\theta$-function inside $\do{R}$ there is no $q'$ dependence inside the Green's functions, and the integral leads to the overall factor:  
\be
2 \alpha' Z k^{2+D} N_D A_D.
\ee
It is easy to see that the remaining $\omega'$-integral, after having taken the derivative in 
$\omega$ and having set $\omega=0$, is of the form:
\be 
\int \frac{d \omega'}{2 \pi i} \frac{1}{D(\omega',0)^4} Tr\Big[O_+ N(\omega',q') \Gamma^{(2)}_{k \psidag} N(\omega',q')O_-N(\omega',q')\Gamma^{(2)}_{k \psi} N(\omega',q')  \Big],
\label{trace-eta}
\ee
where we have used the abbreviations
\ba
O_{\pm}= \left( \begin{array}{cc} 0 & 1 \\  \pm 1& 0
\end{array} \right).
\ea
The $\omega'$-integral is easily performed by picking up the residue of the pole in the upper half complex $\omega'$-plane.
The result is a function of the parameters of the potential and depends on the truncation. 
For illustration we only present the simplest case (cubic truncation and expansion around zero fields):     
\be
\eta_k  =  -  2  N_D A_D(\eta_k,\zeta_k) \frac{\tilde{\lambda}^2}{(1 -\tilde{\mu_k} )^3}.
\ee
The general case (higher truncation, general point $(\psi_0,\psidag_0)$ of expansion) is pretty lengthy. For the full result see Appendix C.
   
The derivative with respect to $q^2$ requires some more considerations. The derivatives 
act on the $\theta$-functions inside the regulator function $h_k$ of the  the Green's function  $G_k(\omega+\omega',q+q')$. 
Putting, for simplicity, the two-dimensional 
vector $\bq=(q_x,0)$ and using the relation $\frac{\partial}{\partial (q_x^2)}|_{q_x=0} =
\frac{1}{2} \frac{\partial^2}{(\partial q_x)^2}|_{q_x=0}$ we find that in the 
derivative 
\be
\frac{\partial^2}{(\partial q_x)^2} G(\omega',q+q') =
2{h'}^2 G  O_{+} G  O_{+} G - h'' G  O_{+} G,
\ee
after setting $q_x=0$, only the second term contributes. After some algebra the 
integral over $q'$ leads to a simple factor $k^2$. The analogue of (\ref{trace-eta}) is:
\be 
\int \frac{d \omega'}{2 \pi i} \frac{1}{D(\omega',0)^4} Tr\Big[O_+ N(\omega',q') \Gamma^{(2)}_{k \psidag} N(\omega',q')O_{+}N(\omega',q')\Gamma^{(2)}_{k \psi} N(\omega',q')  \Big].
\ee
The trace expression depends upon the truncation and has to obtained using Mathematica.
Here we only present the result for the cubic truncation, expanded around the origin:
\be
-\eta - \zeta = 
 \frac{N_D }{  D } 
\frac{ \tilde{\lambda}^2 }{(1 -\tilde{\mu}_k )^3}\,,
\ee
while the general expression can be found in Appendix C.
Let us stress that in this approximation the values of the anomalous dimensions 
depend on the field configuration of our polynomial expansion.
As a consequence, the estimate one obtains is not particularly good.
In order to improve the analysis one should go to the full next order of the derivative expansion with field dependent $Z$ and $\alpha'$ appearing in the kinetic terms.
In this case one would need to solve the coupled PDE flow equations of $V$, $Z$ and $\alpha'$, with the anomalous dimensions being considered as spectral parameters of the nonlinear problem with suitable cutoff operators for the fixed point equations.

%%%%%%%%%%%%%%%%%%%%%
\subsection{Numerical results}
%%%%%%%%%%%%%%%%%%%%%

Let us now turn to the numerical results of our search for fixed points.
We return to (\ref{fp-equation}) and insert our results for the anomalous dimensions 
derived in the previous subsection. With these new expression we repeat the calculations 
described in section 3.    
We first study a polynomial expansion around the origin. For cubic and quartic truncations 
we have found a fixed point solution with the familiar stability properties (one negative eigenvalue), 
but with suspiciously large values of the anomalous dimensions.     
Starting from the quintic truncation, the fixed point solution disappeared.
Based upon our experience from the case of zero anomalous dimensions
where the expansion around the origin was found to be slow, we interpret this again as evidence of bad 
convergence, and we dismiss the  results of the cubic and quartic truncation.  

Next we turn to our second approach, the polynomial expansion around a stationary 
point $u_0$ on the $\psi$ axis. Proceeding in exactly the same manner as before, we find 
a fixed point with the familiar stability properties which is robust against changing the 
order of truncation. Moroever, the sequence of truncations shows a good convergence, and the numerical values of the anomalous dimensions are small. Results are shown in Table 3:       \\
\begin{center}
\begin{tabular}{|l|c|c|c|c|c|c|}\hline
\text{truncation} &3&4&5&6&7&8\\ \hline
\text{exponent} $\nu$ & 0.660&0.659&0.635&0.633&0.634&0.634\\ \hline
\text{mass $\tilde{\mu}_{eff}$} &0.356&0.376&0.396&0.396&0.414&0.414
\\ \hline
\text{anom.dim. $ \eta$}&-0.054&-0.087&-0.081&-0.080&-0.080&-0.080
\\ \hline
\text{anom.dim. $ \zeta$}&0.061&0.113&0.118&0.117&0.117&0.117
\\ \hline
\text{ $i\psi_{0,diag}$}&0.0593&0.0754&0.0766&0.0763&0.0762&0.0763
\\ \hline
\text{ $iu_0$}&0.178&0.222&0.226&0.225&0.225&0.225
\\ \hline
\end{tabular} \\ \vspace{0.5cm}
Table 3: Polynomial expansion around $(r,u))=(0,u_0)$.\\
Parameters of the fixed point for different truncations.
\end{center}   
A comparison with Table 2 shows, for the parameters of the potential, a strong similarity:
the effective mass $\mu$ (0.414 vs. 0.397), the position of the stationary point on the 
diagonal (0.0763 vs.0.074) and on the axis (0.225 vs 0.218). 
The quantity which changes more is the critical exponent $\nu$, which seems to become worse, if we compare it to numerical Monte Carlo estimates, which where very close to the asymptotic values of table I and II.
The reason for this should be found in the poorness of the anomalous dimension estimates, which strongly depends here on the field configuration chosen.

In Fig.~\ref{fig4} we show, for comparison, the same flow diagram as in Fig.\ref{cubic}a for the cubic 
truncation with nonvanishing anomalous dimensions:
\begin{figure}
\begin{center}
\includegraphics[width=5cm,height=5cm]{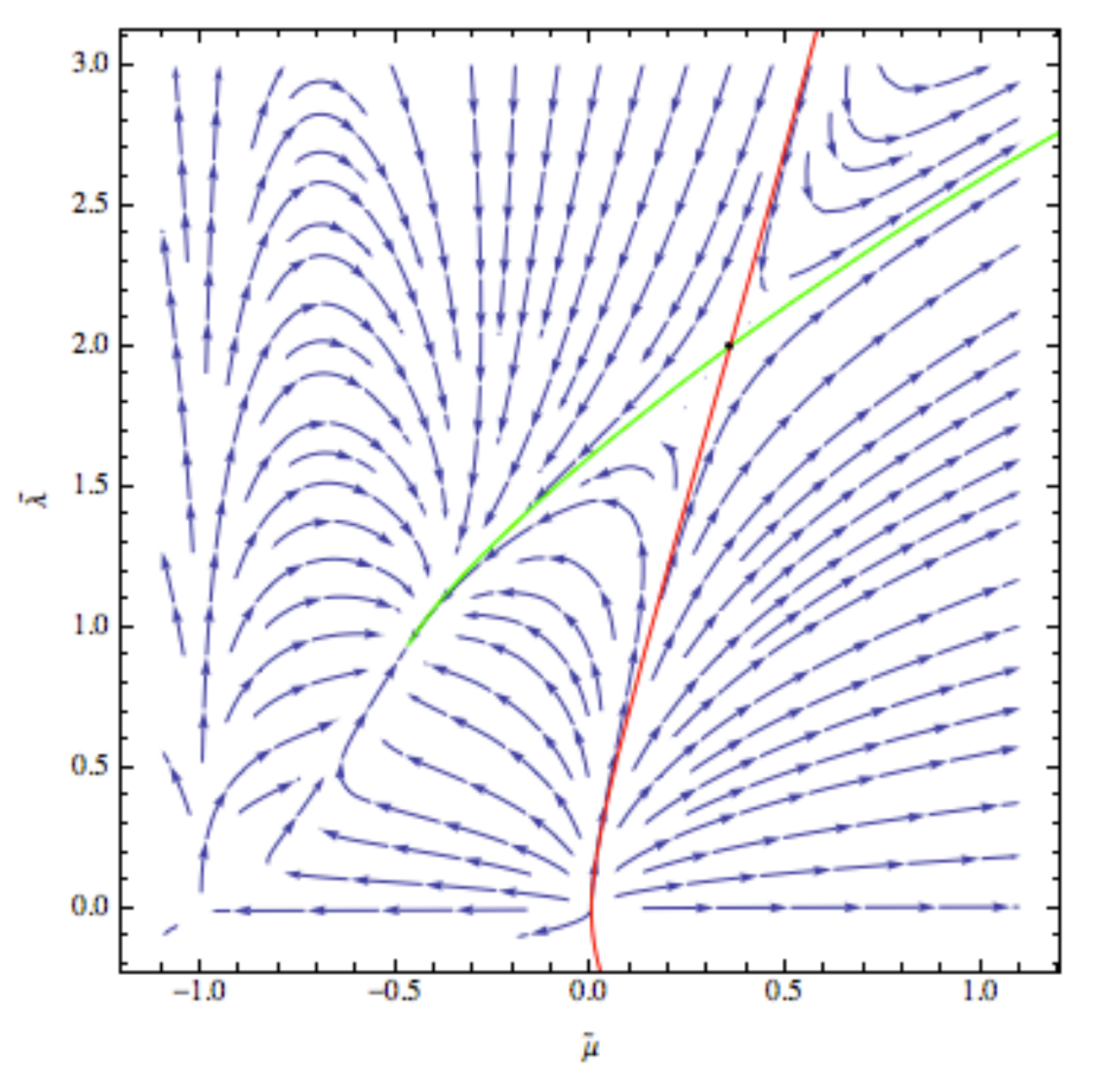}
\hspace{2cm}\\
\caption{Flow in a cubic truncation with nonzero anomalous dimensions.}
\label{fig4}
\end{center}
\end{figure}
Compared to the case without dimensions, the main difference is the appearance of a second (infrared attractive) fixed point:
on the relevant direction (green) to the left of our fixed point. Instead, in Fig.\ref{cubic}a there was a singularity at $\tilde{\mu}=-1$.
Both points change if we move to higher truncations: we therefore interpret them as artifacts of a specific truncation.  
In particular, if we use the polynomial expansion around a stationary point on the axis, Eq.~(\ref{Vaxis}), we find for the vector field of the flow that, with increasing order     
$N$ of the polynomial. the singularity in the variable $\mu_{eff}$ of Eq.~(\ref{effective}),
moves further and further to the left, $\mu_{eff} = -(N-1)$. 
We therefore expect it to disappear in a full analysis without a finite polynomial truncation.
On the whole, our results confirm our expectation that the presence of nonzero anomalous dimensions only leads to quantitative modifications of the fixed-point potential. 

Finally we just shortly comment on an alternative scheme, which we name LPA'${}_2$:
in this scheme the anomalous dimensions are estimated using a configuration of the fields 
at the extremum of the potential along the diagonal ($\tilde\psi=\tilde{\psi}^\dagger$), 
still using a polynomial expansion around the stationary point on the axes to compute the potential. 
We do this for each order of the polynomial by computing iteratively the fixed point potential for a sequence of anomalous dimensions
obtained from the previous step. We find that the convergence of the iterative procedure is very fast and the results are stable with increasing the order of the polynomial. 
In this scheme we find critical exponents $\nu=0.771$, $\eta=-0.263$, $\zeta=0.089$. 
The shape of the fixed point potential is just slightly modified,  with $i u_0=0.274$ and $i \psi_{0,diag}=0.0938$.

In conclusion in the simple LPA' schemes results for the anomalous diemensions depend on the field configuration, and it is needed to improve the determination of the anomalous dimensions along the line previously discussed.

%%%%%%%%%%%%%%%%%%%%%%%%%%%%%%%%%%
\section{A first glimpse at physical interpretations}
%%%%%%%%%%%%%%%%%%%%%%%%%%%%%%%%%%

%%%%%%%%%%%%%%%%%%%%%%%%
\subsection{Flow of physical parameters}
%%%%%%%%%%%%%%%%%%%%%%%%

So far our analysis has been devoted to the search for a fixed point potential; for convenience this has been done in terms of dimensionless parameters. 
In order to make contact with cross sections we need to translate from our dimensionless to physical parameters. 
In order to distinguish between 'dimensionless' and 'dimensionful' quantitites we return to the tilde-notation introduced in section 2. 
Returning to this notation, we have been studying 
the dimensionless fixed point potential $\tilde{V}$ as a function of the dimensionless fields
$\tilde\psi$ and $\tilde{\psidag}$.

Let us first make a few general comments. 
The dimensionful potential in the infrared region is given by
the limit $k \to 0$ of
\be
V_k(\psi,\psidag) = \alpha_k k^{D+2} \tilde{V}_k( Z_k^{1/2} k^{-D/2}\psi,   
 Z_k^{1/2} k^{-D|2}\psidag ),
\label{large-field-behavior}
\ee
i.e. for nonvanishing finite fields $\psi$ and $\psidag$ we are probing the large-field behavior in 
$\tilde{\psi}$ and $\tilde{\psidag}$ of the potential $\tilde{V}_k$. 

We consider a scenario where the bare action for the RFT lies on the critical surface of the non trivial fixed point, 
and for small $k$ we are close to the fixed point.
The general asymptotic behavior (for vanishing anomalous dimensions) is described in eqs.(\ref{large-field-behavior}). 
In the subspace of imaginary fields $\tilde{\psi}$ and $\tilde{\psidag}$  
we have verified that, along the axis and along the diagonal, the fixed point potential indeed approaches this asymptotic behavior.
Generalizing to D dimensions and nonzero anomalous dimensions the canonical power changes to:
\be
\left(\tilde{\psi} \tilde{\psidag} \right)^2 \rightarrow \left(\tilde{\psi} \tilde{\psidag} \right)^{\frac{D+2-\zeta}{D+\eta}}
\ee
In terms of dimensionful quantities and with $Z_k \sim k^{-\eta}$ and $\alpha'_k \sim k^{-\zeta}$, 
(\ref{large-field-behavior}) translates into the small-$k$ behavior 
\be 
\label{quartic}
V_k(\psi,\psidag) \sim \left(\psi \psidag \right)^2 
V_{\infty}(\psi/\psidag).
\ee
The potential becomes quartic and has no quadratic (mass) or cubic terms.
This argument, however, does not apply to the region of very small fields
$\psi$, $\psidag$  and $k$ not small enough such that the dimensionless fields $\tilde{\psi}$, $\tilde{\psidag}$  are not large. 
Our analysis in section 3 shows that in this region the potential $\tilde{V}$ has a more complicated structure, e.g. by exhibiting nontrivial stationary points. 
In order to obtain further insight we shall extend our analysis out of the critical surface starting slightly away from the fixed point
and see how the limit $k\to 0$ is reached. 

It should be noted that in section 3 we also found directions in which the asymptotic behavior is of the form (\ref{large-field-special}):
\be 
\label{quadratic}
V_k(\psi,\psidag) \sim \alpha_k Z_k k^{2} \left(\psi \psidag \right).
\ee
In these directions the potential becomes quadratic and develops a mass: in the limit $k\to 0$, however, this mass goes to zero and the potential becomes a constant.
In our analysis described below we do not find evidence for such a free theory. We therefore conclude that that this asymptotic behavior is exceptional and plays no role     
in our fixed point theory.
 
In the following we will investigate the flow equations for a few dimensionful parameters.    
Of particular interest are the Pomeron intercept which is related to the 'mass' $\mu$:
\be
\alpha(0)-1 = Z_k^{-1} \mu_k =   \alpha'_k k^2 \tilde{\mu}_k,
\ee
and the triple Pomeron vertex which is given by the three-point vertex function:
\be
\lambda_{triple\,\,Pomeron} = Z_k^{-3/2} \lambda_k = \alpha'_k  k^{2-D/2} \tilde{\lambda_k}.
\ee
We also will keep an eye on the Pomeron slope $\alpha'_k$ which determines the 
$t$-slope of the elastic cross section, even if one has to keep in mind that we have not a very reliable estimate of the anomalous dimensions.
The evolution equations for the dimensionless parameters have been discussed before. 
In order to study the flow of the dimensionful parameters, 
we may supplement these equations by the evolution equations of the wave function renormalization $Z(t)$ and of the slope parameter $\alpha'(t)$, 
if we take into account the anomalous dimensions, as defined in (\ref{anomalous-dimensions}).     
    
Let us first return to Fig.\ref{cubic}a. (or Fig.\ref{fig4}),  which - although based on an  inaccurate truncation - 
nevertheless qualitatively correctly illustrates the flow of the (dimensionless) potential parameters. In the following we compute the flow of dimensionful parameters along 
trajectories as shown in Fig.1a. We will distinguish between trajectories inside the critical subspace (in red) which, in the infrared limit, all end at the fixed point, and the flow near  the  relevant direction (in green) for which our fixed point is IR repulsive (UV attractive). This relevant direction has two branches: one (on the upper rhs) goes towards larger values of the effective mass $\tilde{\mu}_{eff}$, the other (on the lower left) goes in the direction of smaller values $\tilde{\mu}_{eff}$.     

Let us begin with the critical subspace. It is spanned by the trajectories which start along
one of the eigenvectors with positive eigenvalues, i.e. in the infrared limit $t \to -\infty$ 
any trajectory in this subspace starting somewhere away from the fixed point will fall into the fixed point. 
As an illustrative example, consider a trajectory on the red line in Fig.\ref{fig4}
starting below the fixed point at $(\tilde{\mu}_{eff},\tilde{\lambda})=(0.061, 0.51)$.   
In Fig.~\ref{fig5} we show, as a function of evolution time $t$, the behavior of the effective mass and of the dimensionful triple coupling:
both parameters vanish in the infrared  limit. 
As discussed in (\ref{quartic}), this indicates that we are ending in an interacting massless phase where the potential contains only the quartic interaction. 
We have verified that these findings remain valid also in higher truncations.
\begin{figure}
\begin{center}
\includegraphics[width=5cm,height=5cm]{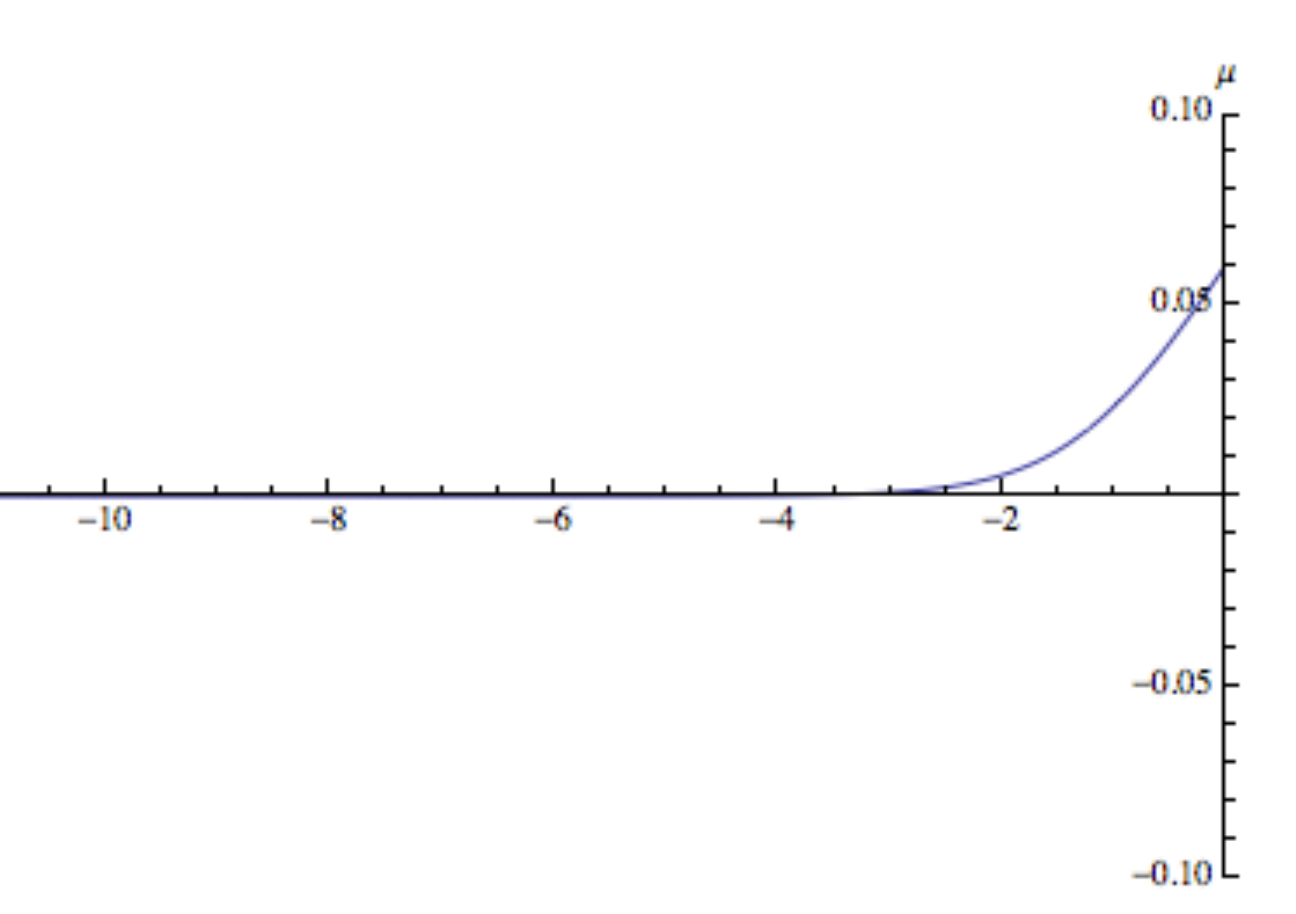}\hspace{2cm}
\includegraphics[width=5cm,height=5cm]{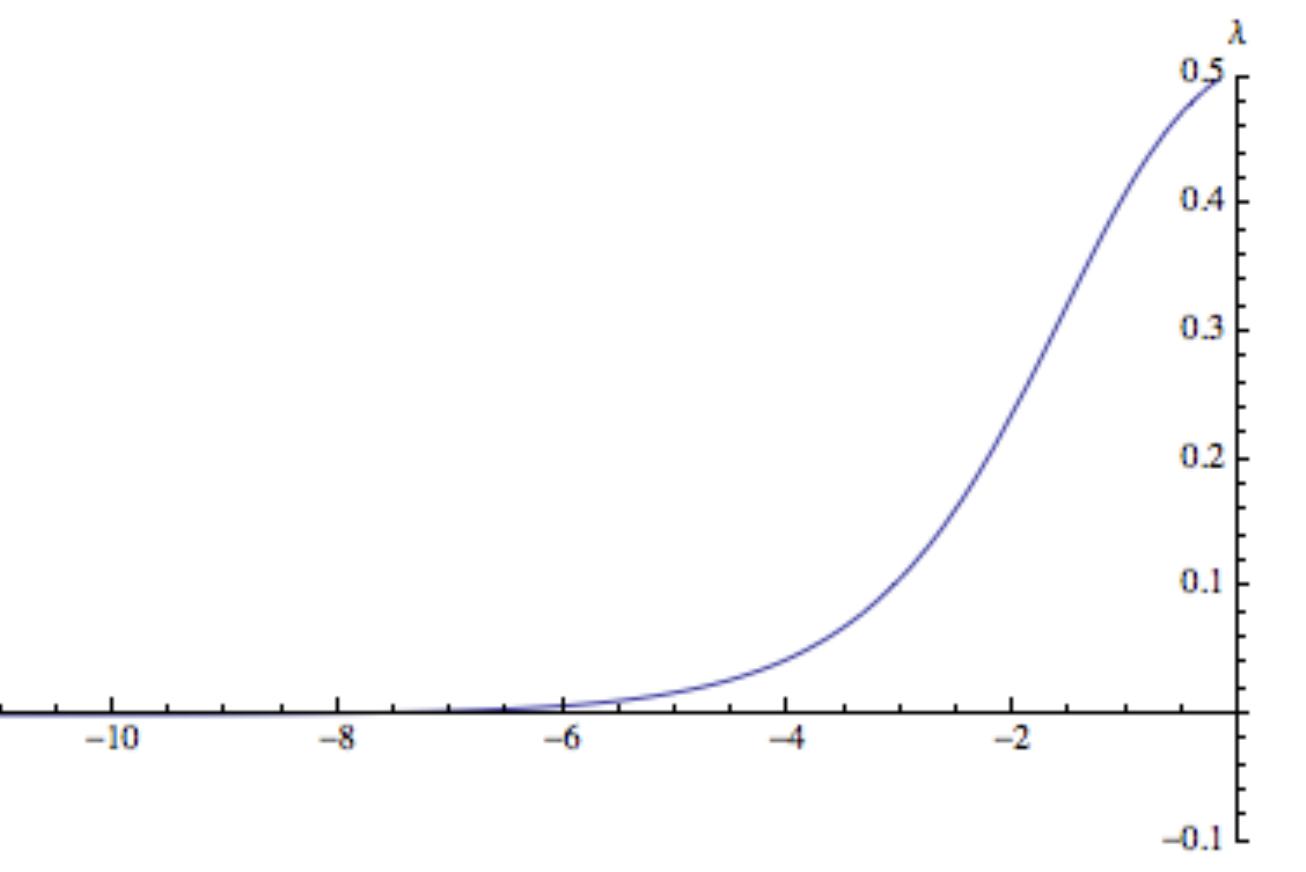}
\caption{Flow of dimensionful parameters (with nonzero anomalous dimensions) inside the critical subspace (red line in Fig.4):
effective mass (left), triple coupling (right).}
\label{fig5}
\end{center}
\end{figure}

Next we consider trajectories which start outside the critical subspace and, in the infrared limit, approach one of the relevant directions.
For this part of our discussion we find it more useful to use a higher, quartic, truncation. 
Again we use the polynomial expansion around the stationary point on the axes, 
which provided a faster convergence in the determination of the fixed point properties.
In Fig.\ref{fig6}a we show the projection of the 4-dimensional parameter space on the two-dimensional $\tilde{\mu}_{eff}$-$\tilde{\lambda}_{eff}$ plane. 
As before, the green line represent the two branches of the relevant direction,
the short red lines the intersection of the critical subspace at the fixpoint values of $\tilde{\lambda}_{1,2}$ and $\tilde{\lambda}_{2,0}$.
\begin{figure}
\begin{center}
\includegraphics[width=6cm,height=6cm]{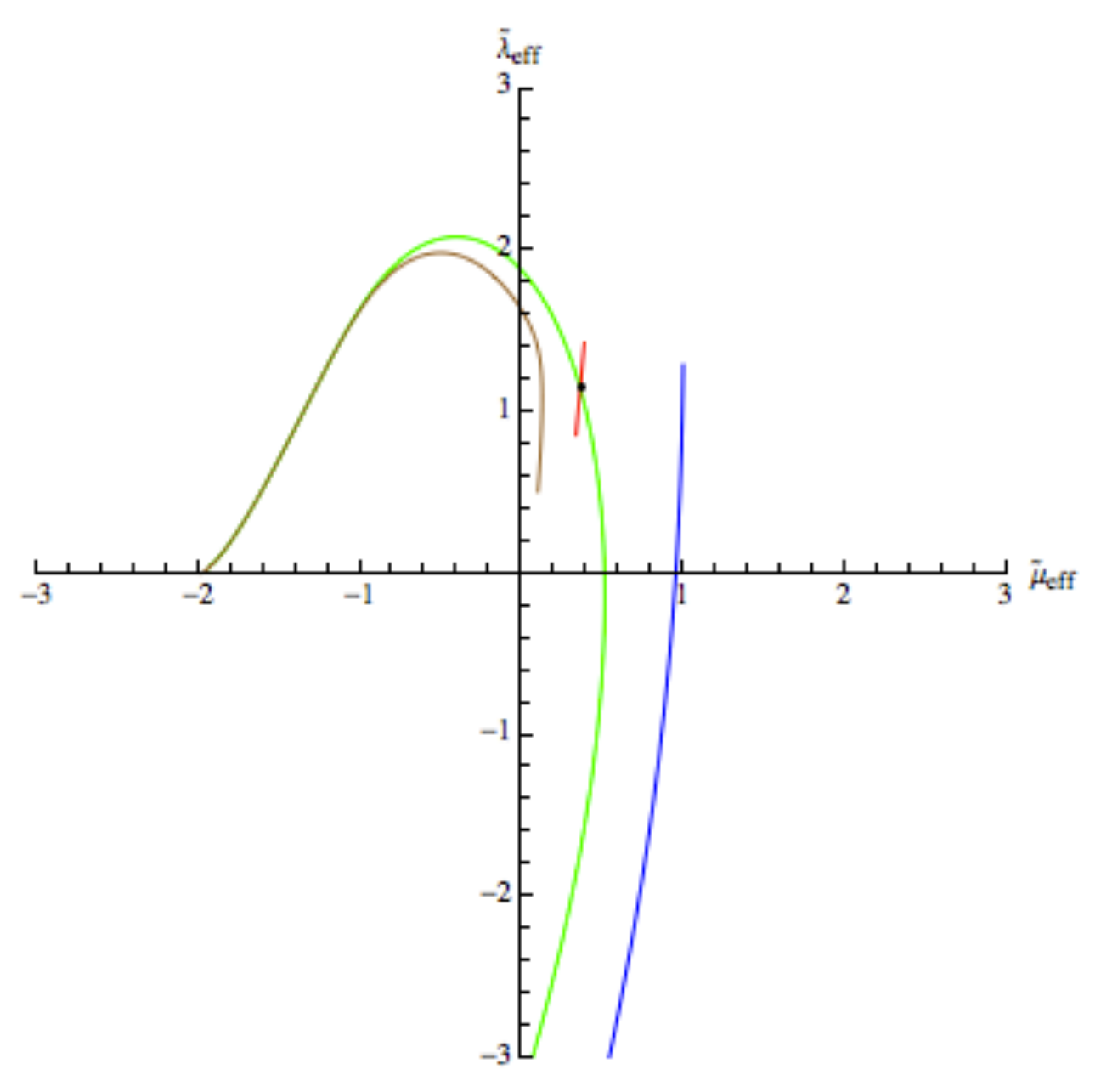}\hspace{1cm}
\includegraphics[width=5cm,height=5cm]{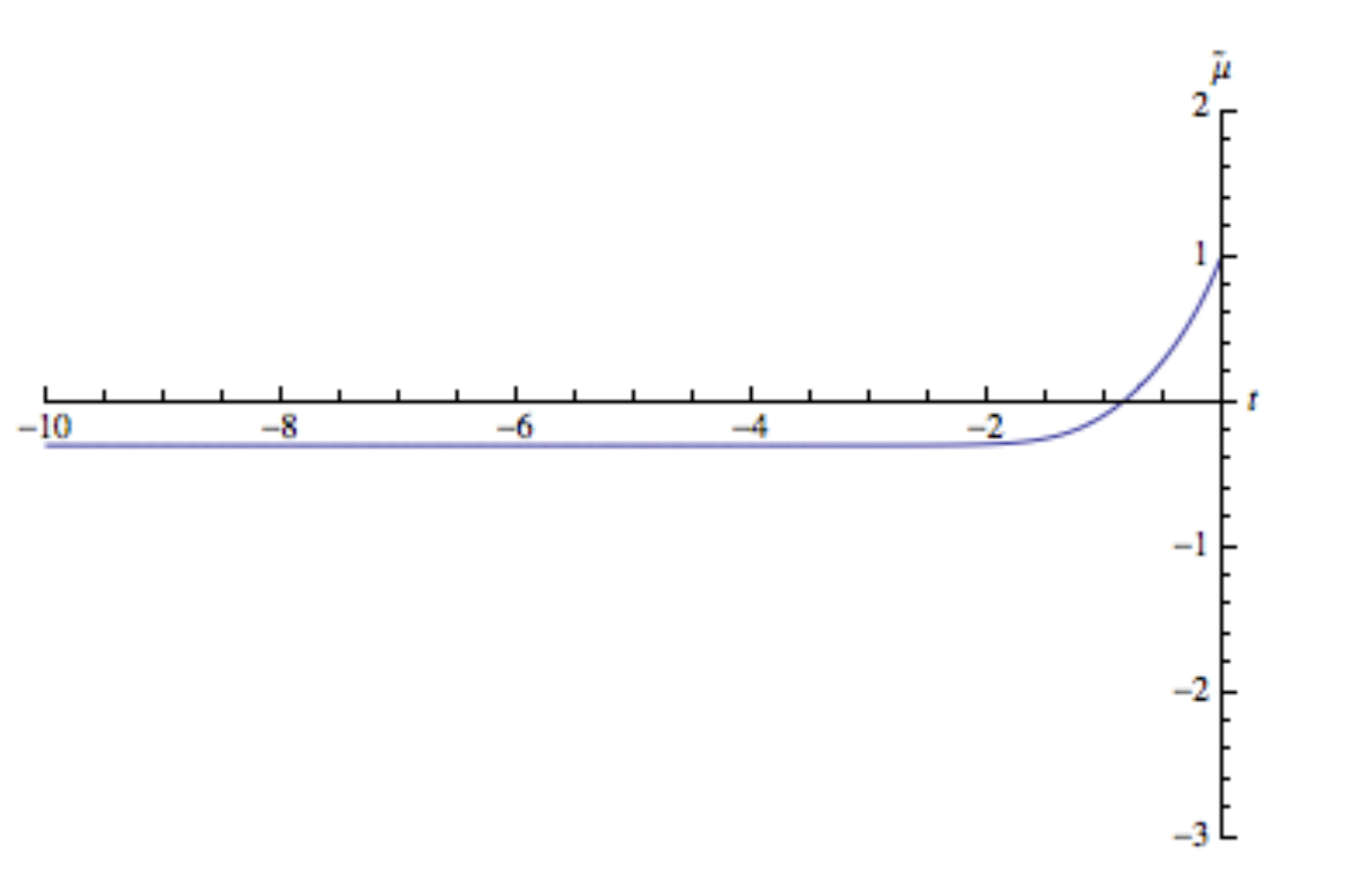}
\caption{ (a) The relevant direction (green), piece of the critical subspace (red), example of two trajectories (blue and brown);
(b) flow of the physical reggeon mass along the blue trajectory.   }
\label{fig6}
\end{center}
\end{figure}
First we note that, when starting exactly on one of branches of the relevant directions,
the dimensionful mass approaches zero. This changes if we chose our starting point somewhat outside the relevant direction (blue line or brown line): near the branch on the rhs the dimensionful mass approaches a negative constant value (Fig.\ref{fig6}b). Also all other dimensionful couplings approach a constant value.

The other branch of the relevant direction (on the left), depending on the order of truncation may end at some singularity
or another fixed point (in our example at $\tilde{\mu}_{eff}=-2$). 
Trajectories starting in the vicinity of this branch, first develop a negative physical mass and 
than also fall into the singularity and end up with zero mass. 
As already discussed in the end of Section $4.2$ we interpret these singularities (which are not stable
under changing the truncation and move to $-\infty$ sending the order of the polynomial truncation to $\infty$) as artifacts connected with the specific truncation and trust the flow only as 
long as the trajectories have not reached the singularity. As a general picture, trajectories which start outside the 
critical subspace (not too far from the relevant direction) approach finite nonzero values of the dimensionful mass
and triple coupling. We find both negative and positive values for the physical mass.  This can interpreted as ending up in an interacting massive phase.   

Let us summarize the numerical results of this subsection as follows. 
The $(n-1)$-dimensional critical subspace divides the $n$-dimensional parameter space into two half-spaces. 
For trajectories inside the critical subspace masses and the triple vertex go to zero as we approach the infrared limit. 
Outside the critical subspace we find, on both sides, trajectories along which masses and couplings approach nonzero limiting  values. 
We interpret these results as finding different phases: 
outside the critical subspace we have massive phases - 'subcritical' for negative masses or 'supercritical' for positive masses - , 
whereas inside the ciritical subspace we approach a massless phase.        

Clearly more sofisticated analysis which include the study of the non perturbative flow of the full 2-point functions in a less crude approximation would be important. This object, non universal, contains important physical informations.
%%%%%%%%%%%%%%%%%%%
\subsection{Possible scenarios}
%%%%%%%%%%%%%%%%%%%

Finally we want to say a few words on an attempt of connecting our analysis with physical cross sections. Our main attention, so far, has been given to the infrared limit $k \to 0$. Depending on  where we start our flow, we either approach the 
relevant direction, with some nonzero mass; alternatively, if we start inside the critical 
subspace we end at the fixed point, with vanishing mass and triple coupling.
How can we translate this into the high energy behavior of cross sections?          

Let us begin with the role of the infrared cutoff, $k$. Our regulator in (\ref{h-regulator}) introduces a 
$k$-dependent reggeon mass and thus suppresses the constribution of small transverse momenta $|\bq|<k$ fluctuations:
\be
\alpha_k(t=-\bq^2)-1 =Z^{-1} \mu_k -\alpha'_k \left( \theta(k^2-q^2) k^2 + \theta(q^2-k^2) q^2\right).
\ee
Equivalently, it suppresses large transverse distances. 

Through the dispersion relation $\omega=\alpha_k (t)-1$ this momentum cutoff constitutes also a cutoff of small values of the "energy" $\omega$. 
Initially the introduction of this regulator was motivated by the singular infrared region in (massless) reggeon field. There might, however, exist some correspondence with features seen in high energy scattering. When comparing 
the high energy behavior of different scattering processes ($\gamma^*\gamma^*$, $\gamma^*p$, $pp$) one observes some systematics: the 'harder' the participating projectiles, the stronger the rise with energy. 'Hardness'
of the projectile is connected  with a 'small extension in the transverse direction': the most prominent example is seen in the virtual photon $\gamma^*$ (virtuality $Q^2$) which fluctuates into a small quark-antiquark system (transverse size $\sim1/Q^2$). Following this line of arguments, one might identify the trend: small transverse 
size - strong rise with energy (large intercept) vs. large size - lower intercep closer to one. In this scheme        
$pp$ scattering has the largest transverse size. The common estimate is
\be
R^2_{pp} = 2 R_p^2 + \alpha' \ln s.
\ee       
At any finite energy (LHC and beyond) the transverse extension of the $pp$ system is far from being infinite.
One might therefore conclude  that at present energies we have not reached the asymptotic region,  
and a QCD-based theoretical description should require some sort of 'finite size' effects. 
In our framework the presence of the infrared cutoff could represent a step in this direction: at present energies a theoretical description with small but finite $k$ applies. 
With increasing energies $k$ becomes smaller and smaller, and only at really 
asymptotic energies we reach the infrared limit. As to the energy dependence 
of $k$, the most simple guess would be logarithmic, i.e. $k\sim \ln s$.  

Applying this interpretation to the flow analysis described above, we see the following possibilities. 
For RG evolution time $t=0$ (which corresponds to some reference scale $k=k_0$) we define a starting value in the space of dimensionless parameters.
If for simplicity we define, as starting values for $Z(t)$ and $\alpha'(t)$, $Z(0)=1$ and $\alpha'(0)=1$,
the dimensionless and dimensionful parameters coincide (numerically for $k_0=1$ in appropriate units). 
Most interesting for us, the starting value for the mass
$\mu$, should be small and positive. We then follow, as a function of evolution time $t$, the flow of the
dimensionless parameters and, simultaneously, the flow of the dimensionful (physical) parameters.   
            
The general case belongs to a trajectory which starts at some arbitrary positive mass outside the critical subspace
(in our analysis we will stay close to the fixed point).
 With decreasing $k$, the trajectory of the dimensionless parameters approaches the relevant direction.
As we have discussed before, the flow of the dimensionful mass $\mu=\alpha(0)-1$ is sensitive to the choice of 
the starting value, but in general  $\mu$ approaches a nonzero negative value. 
At $k\to 0$, this would correspond to an intercept below 1 and thus to falling total cross section 
(with a  power of energy) with finite non zero couplings in the IR limit. 
Alternatively we consider trajectories which start inside the critical surface.    
For these solutions the infrared limit of the dimensionless parameters is the fixed point, and we have seen 
that both the dimensionful mass and the triple coupling vanish:     
this a reggeon field theory with intercept exactly at one and vanishing triple vertex. 
Such a critical theory is not a free theory but has quartic interactions. Since the dimension of the critical subspace is $n-1$ and 
hence less than the dimension $n$ of the full parameter space (the dimension $n$ depends upon the truncation, 
and in the general case will be very large). Such a starting point defines to an exceptional class of solutions.           
   
%%%%%%%%%%%%%%%%%%%%%%%%        
\section{Conclusions}
%%%%%%%%%%%%%%%%%%%%%%%%

In this paper we have addressed the question  whether, in the limit of very high energies and small transverse momenta,
reggeon field theory might provide a theoretical description for high energy scattering.
We have used the functional renormalization group techniques to investigating in a general nonperturbative setting 
and within some approximations (truncation of the theory space)
the RG flow of reggeon field theories in two spatial transverse dimensions as a function of an infrared cutoff, $k$.
As always, an important piece of informations can be obtained looking at the stationary point of the flow, 
since close to criticality one can obtain a certain amount of universal information of the theory, 
so that we have searched for fixed point in a space of possible (local) field theories. 
We have found a candidate which is robust against changing truncations for polynomial expansions of the potential and,
in certain regions of field variables, after a partial expansion, also solves set of coupled differential flow equations.
The fixed point potential is not a universal quantity but its critical exponents are so.
Our result have been obtained for a specific coarse graining scheme (based on a flat optimized regulator).
One should therefore study more general cutoff operators to trace the dependencies.
Nevetheless we find numerical estimates in reasonable agreement with Monte Carlo results coming from studies on directed percolation theory,
which is a dynamical  process belonging to the same universality class of RFT. 

In appendix D we have presented some commeon universal quantities studied in the Monte Carlo, and in the comparison we have also looked at the results given by the two loop perturbation theory in the expansion in $D=4-\epsilon$ transverse dimensions.
While the perturbative expansion may perform well in the determination of the critical exponents we do not expect it to be very useful in predicting the non universal quantities which are caracterized by the transverse space dimensionality $D=2$. 
Indeed in such dimension, e.g. staying in a polynomial truncation approximation for the bare action, the quartic couplings are marginal (while irrelevant in $D=4$). 
Therefore most of the (non universal) features encoded in the emerging 2+1 reggeon field theory arising from QCD which manifest in the flow to the IR
cannot be computed in a reliable way with a perturbative approach in $D=4-\epsilon$. 
We plan to come back to this issue in the future. It will also be important to analyse the connection beween these scaling solutions to reggeon field theory and the results obtained 
by Gribov and Migdal \cite{Gribov:1968fg,Gribov:1968uy}  prior to the renormalization group solutions.   

We then have started to investigate the flow as a function of the 
infrared cutoff, in particular in the limit $k \to 0$. 
We have done this using a simple polynomial expansion with increasing orders, 
but a more refined numerical analysis based on the full solution of the PDE associated to the flow should be considered as a necessary next step.
In the $n$-dimensional space of the parameters of the 
effective potential (here $n$  depends upon the truncation used for the numerical study, and it can be large)        
we found a $n-1$-dimensional 'critical' subspace which divides the full parameter space into two half-spaces:
inside this subspace all trajectories end at the infrared fixed point, where the reggeon mass and the triple 
Pomeron coupling vanish and only quartic interactions remain (massless phase). 
Orthogonal to this $n-1$-dimensional subspace there exists one relevant direction for which the fixed point is repulsive in the infrared limit. 
As expected, trajectories which start outside (but not too far from) the critical subspace, for small $k$, are attracted by this relevant direction. 
For such trajectories the physical (i.e. dimensionful) reggeon mass 
$\mu=Z (\alpha(0)-1)$
has been found to approach constant (often: negative) values (massive phase). 
As a result, it is the choice of the starting value of the flow which determines in which phase the infrared limit is located.
We stress that it is the fundamental theory of strong interaction (QCD)
which should eventually tell at which scale and from which point in theory space
the "bare" effective reggeon field theory (whose fields emerge as composite structures in terms of the QCD fields)
can be considered a possible convenient description so that we can study its flow into the IR. 

We should also note that QCD is a unitary theory. By construction, RFT satisfies unitarity in the 
$t$-channel; whether it also satisifies $s$-channel unitarity is  {\it a priori} not clear and needs  to be studied separately\footnote{Unitarity of the QCD S-matrix should not be confused with the well-known  non-hermiticity of the reggeon field theory hamiltonian.} 
The most obvious constraint follows from the Froissart bound: the requirement for a "bare" reggeon field theory to be compatible with unitarity
implies that its flow ends with a dimensionful intercept not greater than one.
This allows for both a critical reggeon theory, with a "bare" RFT action on the critical surface, 
and also a "bare" action emerging in the massive phase with negative mass.

This theory contains nonperturbative parameters, most notable the Pomeron 
slope $\alpha'$ which enters the transverse growth of the system of scattering projectiles (in the picture of 
Feynmans  'wee partons', it governs the diffusion of 'wee partons').  
For such an interpretation it is important to note that, at present (large but finite) energies the transverse size of scattering systems
(e.g. $pp$ scattering at the LHC) is far from being infinite, and a successful theoretical description might very well contain 
some infrared cutoff in the transverse direction, which goes to zero as energy goes to infinity. A possible theoretical picture 
could consist of a flow of reggeon field theories: in the UV region one starts with QCD reggeon field theory (BFKL),
still obtainable with perturbative investigations, and taking smaller and smaller values of the cutoff parameter $k$ 
one eventually will end at a IR behavior as we have obtained in this paper. 

We have not yet started to look into any phenomenological application. 
We find it appealing to see the possiblity that a start (at finite $k$) with a positive reggeon mass (Pomeron intercept above one)
can lead to an infrared limit where the intercept is exactly at one: a rising total cross section, accompanied by a finite transverse
size, will eventually turn into a universal constant (modulo possible logarithms) cross section.  
This is the critical scenario, which, in order to be realized, would need a non trivial constraint for the "bare" RFT in the theory space,
since it should lie exactly on the critical surface. 
One may wonder if unitarity in QCD could play such a role.

The other more generic scenario, a massive phase with a "bare action" not belonging to the critical surface,
can nevertheless approximate arbitrarily well an intercept equal one  in the IR limit.
We stress that both the scenarios would are related to interacting theories in the IR limit, with the critical theory having a much simpler structure.
 
This clearly opens a large field of questions which have to be studied in detail, both refinements of our theoretical 
analysis and closer looks into phenomenology.
A natural extension of the effective Regge limit dynamics in QCD can be constructed in order to take into account the odderon reggeon
which is the parity odd partner of the pomeron, encoding for example the difference
in cross sections of $pp$ and $p\bar{p}$ scattering process~\cite{Lukaszuk:1973nt}.
At the perturbative level the Odderon in QCD appears as a composite object of three reggeized gluons in the fundamental representation, and in contrast to the BFKL Pomeron, 
it has the remarkable property of having the intercept at l~\cite{Bartels:1999yt,Bartels:2013yga}.
Some perturbative interaction vertices among pomeron and odderon states have been studied in ~\cite{Bartels:1999aw, Bartels:2004hb}.
The effective action of such a pomeron-odderon model can be  constructed by symmetry requirements and analyzed nonperturbatively with the same functional renormalization group methods as used in this paper. 
An interesting question is again related to the universal behavior of this extendend model, 
which should have a counterpart in reaction-diffusion non equilibrium phenomena.

Needless to say that having at our disposal a possible approximate simple effective description of  QCD phenomena in the Regge limit would add
insights in our understanding of strong interaction.    \\ \\

%%%%%%%%%%%%%%%%%%%%%%%%%%%%%%%%%%%%%%%%%%%%%%%%%%%%%%
\noindent           
{\bf Acknowledgements:}
This work has been supported by the grant MEC - 80140070, Fondecyt: 1120360 and  1140842 and  DGIP 11.15.41. Most of these investigations have been performed
at the Departamento de Fisica, Universidad Tecnica Federico Santa Maria (UTFSM), Valparaiso, Chile. 
J.B. also gratefully acknowlegdes the support of the INFN and the hospitality of the Bologna University,
C.C. thanks for the hospitality and support of the Bologna and the Hamburg University .
G.P.V. is grateful for hospitality and support of the UTFSM and the Hamburg University.

%%%%%%%%%%%%%%%%%%%%%%%%%%%%%%%%%%%%%%%%%%%%%%%%%%%%%%
\appendix
%%%%%%%%%%%%%%%%%%
\section{$\beta$-functions}
%%%%%%%%%%%%%%%%%%

In this appendix we list, for illustration, the $\beta$-functions for a quartic 
truncation:
\ba
\label{betamu}
\dot{\tilde{\mu}}&=&  \tilde{\mu}  (-2+\zeta +\eta)+ 2N_D A_D(\eta_k,\zeta_k) \frac{ \tilde{\lambda} ^2  }{  ( 1-\tilde{\mu} )^2}, \\
\label{betalambda}
\dot{\tilde{\lambda}}&=& \tilde{\lambda} \left( (-2+\zeta +\frac{D}{2}+\frac{3\eta}{2} )+ 2N_D A_D(\eta_k,\zeta_k) 
\left(\frac{ 4\tilde{\lambda}^2}{  (1-\tilde{\mu} )^3}+ \frac{(\tilde{g}+3\tilde{g}') }{ (1-\tilde{\mu})^2} \right)\right),\\
\label{betag}
\dot{\tilde{g}}&=&   \tilde{g} (-2+D+\zeta +2\eta )+2N_D A_D(\eta_k,\zeta_k) \left( \frac{ 27 \tilde{\lambda} ^4}{  (1-\tilde{\mu} )^4}+ \frac{(16 \tilde{g} + 24 \tilde{g}') \tilde{\lambda} ^2 }{  (1-\tilde{\mu} )^3}+ \frac{ 
 \left(\tilde{g}^2+9 \tilde{g}'^2\right)} {  (1-\tilde{\mu} )^2}\right),\nonumber\\
\\
\label{betagpr}
\dot{\tilde{g}}'&=& \tilde{g}' (-2+D+\zeta  +2\eta )+2N_D A_D(\eta_k,\zeta_k)
 \left( \frac{ 12 \tilde{\lambda} ^4}{  (1-\tilde{\mu} )^4}+ \frac{(4 \tilde{g}+18 \tilde{g}') \tilde{\lambda} ^2 }{  (1-\tilde{\mu} )^3}+ \frac{3
 \tilde{g}\tilde{g}'} {  (1-\tilde{\mu} )^2}\right).\nonumber\\ 
\ea
  
\section{Solving differential equations}
In this appendix we describe in some detail our numerical solutions of the differential 
(\ref{fp-equation}). As described in section 3, we restrict ourselves to the neighbourhod of 
the $\phi$ axis, $\phidag=0$, and of the diagonal line $\phi=\phidag$. Our study of polynomial expansions has lead to a consistent picture of the region small fields, now we 
are interested in the large-field region. In particular, we want to analyze the structure at large fields beyond the stationary points.         
    
We begin with the analysis in the vicinity of the $\psi$-axis, and perform a small-r expansion  (eq.(\ref{diff-axis})): 
\be
V=\phi\phidag f_{a1}{\phi_+} + (\phi\phidag)^2 f_{a2}(\phi_+) + (\phi\phidag)^3 f_{a3}(\phi_+).
\ee
Inserting this ansatz into the differential equation
(\ref{fp-equation}) and expanding in powers of $r$, we obtain a set of three coupled differential equations of second order for the functions $f_{ai}(\phi_+)$, $i=1,2,3$, which  
we want so solve numerically.  These equations are easily obtained using symbolic computational tools like Mathematica and will not be listed here.
To solve these equations we proceed in four steps.\\  
(1) First we fix the initial conditions at zero fields.
To this end we make a polynomial ansatz for the functions $f_{ai}$ (of the order $n_1=17$, $n_2=15$, $n_3=13$ for the functions the functions $f_{a1}(\phi_+)$, $f_{a2}(\phi_+)$, $f_{a3}(\phi_+)$ resp.):
\ba
Pol_1(\phi_+)=\sum_n^{n1} \lambda1_n \phi_+^n \nonumber\\
Pol_2(\phi_+)=\sum_n^{n2} \lambda2_n \phi_+^n \nonumber\\
Pol_3(\phi_+)=\sum_n^{n3} \lambda3_n \phi_+^n .
\ea
Inserting these polynomials into the three differential equations and expanding in powers 
$u$, we obtain a set of algebraic equations for the coefficients $\lambda1_n$, $\lambda2_n$, $\lambda3_n$
which we solve numerically.\\ 
(2) Beginning with the region of negative $\phi$, $\phidag$, we use these  polynomials  
at a matching point with very small, $\phi_+=-0.01$,  as initial conditions for the three differential equations. Because of the strong nonlinearities of the differential equations, 
the numerical solutions turn out to be valid only in a finite region of $\phi_+$, and it is not possible to find solutions which extend to infinitely large values $\phi_+$. \\
(3) We therefore make an analytic ansatz for large negative  fields, using the asymptotic behavior (\ref{large-field-behavior}):
\ba
f_{a1, \infty }(\phi_+)&=&A_1 \phi_+^2 \sum^{nmax} a_n \phi_+^{-n} \nonumber\\
f_{a2,\infty}(\phi_+)&=&A_2 \sum^{nmax} b_n \phi_+^{-n} \nonumber\\
f_{a3,\infty}(\phi_+)&=&A_3 \phi_+^{-2} \sum^ {nmax} c_n \phi_+^{-n} .
\label{asympt-solution}
\ea
After inserting this ansatz into the differential equations, and we arrive at a set of recursion 
relations for the expansion coefficients $a_n,b_n,c_n$, all depending on the three unknown parameters $A_1$, $A_2$, $A_3$. These numerical values  are fixed by matching our asymptotic solution to the numerical solution from step (2): $(A_1,A_2,A_3)=(-4.0,\, -0.754,\, 0.044)$. \\
(4) For positive $\phi_+$-values we again use, as in step (2), the expansion around the orgin as 
initial conditions for the three differential equations.
The matching is done at $\phi_+=0.01$.  In contrast to the negative $\phi_+$-region, the numerical solution is stable up to large negative $\phi_+$,
and the asymptotic behavior is the one shown in  (\ref{large-field-special}).
We summarize these results by showing in Fig.2  the function$f_{a1}(\phi_+)$. Further details are discussed in section 3. 

Next we discuss the vicinity of the diagonal, $\phi=\phidag$. Using the variables
$\phi_+$ and $\phi_-$ (eq.(\ref{variables-pm})), we expand in powers of $\phi_-^2$:
\be
V=f_{d0}(\phi_+) + \phi_-^2 f_{d2}(\phi_+) +\phi_-^4 f_{d4}(\phi_+). 
\ee 
As before, we insert this ansatz into the differential equation
(\ref{fp-equation}), expanding in powers of $\phi_-^2$, and obtain a set of three coupled differential equations of second order for the functions $f_{di}(\phi_+)$, $i=0,2,4$, which  
we have to solve numerically. These equations are lengthy and will not be listed here. To solve these equations we again proceed in four steps.\\
(1) For the initial conditions at zero we make us of the polynomial expansion around the origin, choosing the highest truncation of order 12 (cf. Table 1).\\
(2) Starting at $\phi_+=-0.01$ we solve the differential equations numerically. Again, this 
solution does not extend to large values of $\phi_+$ (up to $\phi_+=-0.195$).\\
(3) For large fields we make an ansatz of the form (\ref{asympt-solution}) with constants 
$B_0,B_2,B_4$; From the differential equations we derive a set of recursion relations for the 
coefficients; the resulting expessions depend upon the three unknown parameters
$B_0,B_2,B_4$. Their values are fixed by matching the asymptotic solution to the numerical 
solution found in step (2): $(B_0,B_2,B_4)= (-17.4,\,-17,\,0.647)$.\\
(4) In the final step we turn to the region of positive $\phi_+$ and proceed in the same way as before: starting at $\phi_+=0.01$
(with initial conditions given by the polynomial expansion around the origin) we numerically solve the differential equation and find that the solution
extends to large positive  values, following the behavior (\ref{large-field-special}).
As discussed in section 3, we show the shape of the potential along the diagonal line:
there is no furter stationary point beyond the one near the origin.            
         
%%%%%%%%%%%
\section{Anomalous dimensions}
%%%%%%%%%%%%%%%%%%
In this appendix we list the explicit expression for the approximated anomalous dimensions for a generic potential 
$\tilde{V}(\tilde \psi^\dagger,\tilde \psi )$, expanded around a general field configuration.
The derivation is described in Section $4.1$. One finds
\ba
\eta=&{}&\frac{1}{4} N_D A_D(\eta,\zeta)
\biggl\{
\frac{3 (v^{(1,1)}+1)}{\left((v^{(1,1)}\!+\!1)^2-v^{(2,0)} v^{(0,2)}\right)^{5/2}} \times \nonumber\\
&{}&
\Bigl[
-(v^{(1,1)}\!+\!1) (v^{(3,0)} v^{(0,3)}-v^{(2,1)} v^{(1,2)})+v^{(0,2)} \left(v^{(3,0)} v^{(1,2)}-(v^{(2,1)})^2\right)\nonumber\\
&{}&
+\frac{(v^{(1,1)}\!+\!1)^2   \left(v^{(2,1)} v^{(0,3)}-(v^{(1,2)})^2\right)}{v^{(0,2)}}
   \Bigr]
   \nonumber\\
&{}&   +\frac{v^{(3,0)} v^{(0,3)}-\frac{3 (v^{(1,1)}\!+\!1) \left(v^{(2,1)} v^{(0,3)}-(v^{(1,2)})^2\right)}{v^{(0,2)}}-v^{(2,1)}
   v^{(1,2)}}{\left((v^{(1,1)}\!+\!1)^2-v^{(2,0)} v^{(0,2)}\right)^{3/2}}
\biggr\} \,,
\ea

\ba
\eta+\zeta=&{}&\frac{N_D}{8 D \left((v^{(1,1)}\!+\!1)^2-v^{(2,0)} v^{(0,2)}\right)^{7/2}} \biggl\{\nonumber\\
&{}&
+v^{(0,3)} v^{(3,0)} \left(2 (v^{(2,0)})^2 (v^{(0,2)})^2+v^{(2,0)} (v^{(1,1)}\!+\!1)^2 v^{(0,2)}+2 (v^{(1,1)}\!+\!1)^4\right)
\nonumber \\
&{}&+5 v^{(0,3)}(v^{(2,0)})^2 (v^{(1,1)}\!+\!1)^2 v^{(1,2)}
-2 (v^{(1,1)}\!+\!1) (v^{(2,1)})^2 v^{(0,2)} \left(3 v^{(2,0)} v^{(0,2)}+2 (v^{(1,1)}\!+\!1)^2\right)\nonumber\\
 &{}&-2 (v^{(1,1)}\!+\!1) v^{(1,2)} \left(3 v^{(2,0)} v^{(0,2)}+2 (v^{(1,1)}\!+\!1)^2\right) (v^{(2,0)} v^{(1,2)}+v^{(3,0)} v^{(0,2)})\nonumber\\
 &{}&+  v^{(2,1)}
  \Biggl( 5 v^{(3,0)} (v^{(1,1)}\!+\!1)^2 (v^{(0,2)})^2
   - 2 v^{(2,0)} (v^{(1,1)}\!+\!1) v^{(0,3)} \left(3 v^{(2,0)} v^{(0,2)}+2 (v^{(1,1)}\!+\!1)^2\right) \nonumber\\
&{}& +v^{(1,2)} \left(6 (v^{(2,0)})^2 (v^{(0,2)})^2+17 v^{(2,0)} (v^{(1,1)}\!+\!1)^2 v^{(0,2)}+2 (v^{(1,1)}\!+\!1)^4\right)
   \Biggr)
\biggl\}\,
\ea
Here $v^{(m,n)}$ denotes derivatives of the potential $\tilde{V}$, evaluated at the 
point in field space, $\tilde{\psi}^\dagger_0$, $\tilde{\psi}_0$, around which we expand:
\be
v^{(m,n)}= \frac{\partial^{m+n}\tilde V}{(\partial \tilde{\psi}^\dagger)^m (\partial \tilde{\psi})^n} |_{\tilde{\psi}^\dagger_0,\tilde{\psi}_0} 
\ee  
For the case of the stationary point being on one of the axis' the expressions simplify considerably.

%%%%%%%%%%%%%%%%%%
\section{Universal quantities and comparison with Monte Carlo and perturbative results}
In this appendix we want to comment on some relations which are valid in reggeon field theory as well as in the directed percolation process.

We consider the anisotropic scaling transformations
\be
x\to \Lambda \,x ,\quad \tau \to  \Lambda^z \tau\, , 
\ee
\be
\psi(\tau,x) \to \Lambda^{-D/2-\eta/2} \psi(\Lambda^z  \tau, \Lambda \,x) ,\quad
\psidag(\tau,x) \to \Lambda^{-D/2-\eta/2} \psidag(\Lambda^z  \tau, \Lambda \,x)\,.
\ee
The anomalous dimensions $z=2-\zeta$ and $\eta$, in their scale dependent form associated to the scale dependent EAA,  were defined in Eq.~(\ref{anomalous-dimensions}).
At the fixed point they become scale independent and $Z$ and $\alpha'$ acquire a pure power law scaling behavior.
It is also convenient to introduce the parameter $\Delta$, which describes the deviation from the criticality as the distance
from the FP along the single relevant direction, with critical exponent $\nu$, which in general is named $\nu_\perp$.
One can view $\Delta$ as the coordinate associated to the relevant deforming operator in the basis of the eigenoperators. 
This means that under rescaling one has $\Delta \to \Lambda^{-\frac{1}{\nu_\perp}} \Delta$.
Then one usually defines $\nu_\parallel=z \nu_\perp$, according to the fact that  close to criticality the correlation lengths in space and time scale
as $\xi_\perp\sim\Delta^{-\nu_\perp}$ and $\xi_\parallel\sim\Delta^{-\nu_\parallel}$, respectively.
In percolation theory one uses $\Delta=p-p_c$ which measures the deviation from the critical probability.

One can construct the following quantities invariant under rescaling: $x \tau^{-\frac{1}{z}}$ and $\Delta \tau^{\frac{1}{\nu_\parallel}}$.
Therefore at criticality the 2-point function can be written as
\be
G(x,\tau) \sim \tau^{\frac{-D-\eta}{z}} F(x \tau^{-\frac{1}{z}},\Delta \tau^{\frac{1}{\nu_\parallel}})\sim |x|^{-D-\eta} F(x \tau^{-\frac{1}{z}},\Delta \tau^{\frac{1}{\nu_\parallel}})
\ee
where the arguments of $F$ are scale invariant, while the function $F$ is not universal. In particular, the quantity $x \tau^{-\frac{1}{z}}$ tells
how at criticality reggeon field theory, which is still interacting, predicts for scale invariant quantities a particular transverse scale-rapidity
dependence dictated by a one dimensional curve in the plane.

There are only three independent critical exponents $\nu=\nu_\perp$, $z$ and $\eta$.
Usually in directed percolation another critical exponent named $\beta$ is commonly considered, clearly dependent on the previous ones.
It associated to the scaling of a density $\rho\sim \Delta^\beta$. It is related to the other critical exponents by the so called hyperscaling relation $\beta=\frac{\nu_\perp}{2}(D+\eta)$.
This exponent is useful when comparing with Monte Carlo results, which usually are represented in terms of $\nu_\perp$, $\nu_\parallel$ (or $z$) and $\beta$.

It may be interesting to compare also with critical exponents obtained in the $\epsilon$-expansion of RFT near $D=4$. We first present a mapping between the critical exponent notation used in this paper and the one used in the RFT at $D=4-\epsilon$ (for the latter ones we put a subscript  'o' for the critical exponents):
\be
z_o=\frac{2}{z}, \quad \eta_o=-\frac{\eta}{z}, \quad \nu_o=\nu_\parallel , \quad \beta_o=\beta.
\ee
The perturbative two loop computation for $D=4-\epsilon$ is available since many years~\cite{Bronzan:1974nw,Janssen}:
\ba
\nu&=&\frac{1}{2}+\frac{\epsilon}{16}+\left(\frac{107}{32}-\frac{17}{16}\ln {\frac{4}{3}} \right) (\epsilon/12)^2+{\cal O}(\epsilon^3) \nonumber\\
z&=&2-\frac{\epsilon}{12} -\left(\frac{67}{24}+\frac{59}{12}\ln {\frac{4}{3}} \right) (\epsilon/12)^2+{\cal O}(\epsilon^3)\nonumber\\
\beta&=&1-\frac{\epsilon}{6}+\left(\frac{11}{12}-\frac{53}{6}\ln {\frac{4}{3}} \right) (\epsilon/12)^2+{\cal O}(\epsilon^3) \nonumber\\
\eta&=&-\frac{\epsilon}{6} -\left(\frac{25}{12}+\frac{161}{6}\ln {\frac{4}{3}} \right) (\epsilon/12)^2+{\cal O}(\epsilon^3)
\ea
We finally compare our findings with the Monte Carlo results which are usually represented in terms of $\nu=\nu_\perp$, $\nu_\parallel$ (or $z$) and $\beta$.

In Table 4 we summarize and compare our results in $D=2$ spatial dimensions, obtained in LPA (from table 2), LP' (from table 3), LPA'$_2$ (from the text at the end of section2), with the two-loop $\epsilon$-expansion
and with recent Monte Carlo data.
\begin{center}
\begin{tabular}{|c|c|c|c|c|c|} \hline
		&  LPA & LPA' & LPA'${}_2$ & 2-loop $\epsilon$-exp. & MC~\cite{MC2}\\ \hline
 $\nu$ 	& 0.730 &	0.634	&0.771	&0.709	&0.729\\ \hline
 $z$		& 2	& 1.883	&1.911	& 1.716	&1.766 \\ \hline
$\beta$	& 0.730	& 0.608 & 0.669 & 0.622 & 0.580\\ \hline
\end{tabular} \\ \vspace{0.5cm}
Table 4: Comparison of our results in the LPA , LPA', and LPA'${}_2$ truncations \\
with the two-loop epsilon expansion and recent Monte Carlo results\\ for directed percolation in $d=2+1$.
\end{center}     
One notices that the determination of the critical exponent $\nu$ is almost exact in the simplest LPA truncation. 
This is a common feature in several FRG studies applied to other theories. In contrast, $\beta$ is better approximated in the LPA' truncation. 
We notice that when taking into account the anomalous dimension in LPA' scheme, 
this truncation reduces the quality of the estimate for $\nu$.
Moreover, we also report the results of LPA'${}_2$ obtained by evaluating the anomalous dimensions
(with an iterating procedure until convergence is reached) on the extremum of the potential on the diagonal). 
This scheme gives a better estimate of the anomalous dimension $\eta$, but a worse one for $\zeta$.
The determination of $z$ is better in the two loop perturbation theory which gives a reasonable good description for the all the universal parameters. Here
one should keep in mind that, although the universal parameters seem to be 
well-approximated by the (perturbative) $\epsilon$-expansion,
one cannot expect a reliable result from perturbation theory in $D=4-\epsilon$ for the non universal quantities which affect physical predictions:  in $D=2$ the number of relevant/marginal operators is greater than in $D=4$. As a result, one needs to employ a 
nonperturbative computation.
In order to improve our nonperturbative estimates one has to go to the full leading order in the derivative expansion, with field dependent $Z$ and $\alpha'$ functions.

Finally, we mention that In our investigation of reggeon field theory we find, in presence of anomalous dimensions (LPA' scheme), sligthly different values for $\nu$ and $\beta$ compared to the ones in~\cite{Canet:2003yu}, where $\nu=0.623$ and $\beta=0.597$ are given.
Since it seems that the same RG scheme is used it is not clear which is the source of the discrepancy.

%%%%%%%%%%%%%%%%%%%%%%%%%%%%%%%%%%%%%%%%%%%%%%%%%%%%%%%%%%
        
\end{document}